%% file: BPH-11-011_temp.tex
\pdfoutput=1

\documentclass[11pt,twoside,a4paper,cmspaper,final,collab]{cms-tdr}
\def\svnVersion{185700}\def\cmsCernNo{2013-014}\def\cmsCernDate{\today}\def\cmsMessage{Submitted to the Journal of High Energy Physics}

\begin{document}\cmsNoteHeader{BPH-11-011}

\hyphenation{had-ron-i-za-tion}
\hyphenation{cal-or-i-me-ter}
\hyphenation{de-vices}

\RCS$Revision: 172153 $
\RCS$HeadURL: svn+ssh://svn.cern.ch/reps/tdr2/papers/BPH-11-011/trunk/BPH-11-011.tex $
\RCS$Id: BPH-11-011.tex 172153 2013-02-15 22:05:01Z alverson $
\newlength\cmsFigWidth
\ifthenelse{\boolean{cms@external}}{\setlength\cmsFigWidth{0.85\columnwidth}}{\setlength\cmsFigWidth{0.4\textwidth}}
\ifthenelse{\boolean{cms@external}}{\providecommand{\cmsLeft}{top}}{\providecommand{\cmsLeft}{left}}
\ifthenelse{\boolean{cms@external}}{\providecommand{\cmsRight}{bottom}}{\providecommand{\cmsRight}{right}}
\newcommand{\Xnew}{\ensuremath{\mathrm{X}(3872)}\xspace}
\newcommand{\psitwo}{\ensuremath{\psi(\mathrm{2S})}\xspace}
\cmsNoteHeader{BPH-11-011} 
\title{\texorpdfstring{Measurement of the \Xnew production cross section via decays to $\JPsi \Pgpp\Pgpm$ in pp collisions at $\sqrt{s} = 7\TeV$}{Measurement of the X(3872) production cross section via decays to J/psi pi pi in pp collisions at sqrt(s) = 7 TeV}}

\date{\today}

\abstract{The production of the $\Xnew$ is studied in pp collisions at $\sqrt{s}=7$\TeV, using decays to $\JPsi \Pgpp \Pgpm$, where the $\JPsi$ decays to two muons. The data were recorded by the CMS experiment and correspond to an integrated luminosity of 4.8\fbinv. The measurements are performed in a kinematic range in which the $\Xnew$ candidates have a transverse momentum $10< \pt < 50$\GeV and rapidity $|y|<1.2$. The ratio of the $\Xnew$ and $\psitwo$ cross sections times their branching fractions into $\JPsi \Pgpp \Pgpm$ is measured as a function of \pt. In addition, the fraction of $\Xnew$ originating from B decays is determined.
From these measurements the prompt $\Xnew$ differential cross section times branching fraction as a function of \pt is extracted.
The $\Pgpp \Pgpm$ mass spectrum of the $\JPsi \Pgpp \Pgpm$ system in the $\Xnew$ decays is also investigated.}

\hypersetup{%
pdfauthor={CMS Collaboration},%
pdftitle={Measurement of the X(3872) production via decays to J/psi pi pi in pp collisions at sqrt(s) = 7 TeV},%
pdfsubject={CMS},%
pdfkeywords={CMS, physics}}

\maketitle 

\section{Introduction\label{sec:intro}}

The discovery of the $\Xnew$ resonance by the Belle experiment in 2003~\cite{bib-Choi:2003ue} and its subsequent confirmation by BaBar, CDF, and \DZERO~\cite{bib-Aubert:2004ns,bib-Acosta:2003zx,bib-Abazov:2004kp} has attracted a large interest in ``exotic'' quarkonium spectroscopy since it was the first observation of an unexpected charmonium candidate.
Many new unconventional states with masses above the open-charm or open-bottom thresholds, $m> m(\mathrm{D}\overline{\mathrm{D}})$ and $m> m(\mathrm{B}\overline{\mathrm{B}})$, respectively, have been observed~\cite{bib-Brambilla:2010cs}.
There are several interpretations of the $\Xnew$ state: a charmonium state, a $\mathrm{D}^*\overline{\mathrm{D}}$ molecule, or a tetraquark state~\cite{bib-Brambilla:2010cs}.
The $\Xnew$ has been observed in several decay channels, including $\JPsi \Pgpp\Pgpm$, $\mathrm{D}^*\overline{\mathrm{D}}$, $\JPsi \gamma$, $\psitwo \gamma$, and $\JPsi \omega $.
The analysis of the $\Xnew$ angular distributions in decays to $\JPsi \Pgpp\Pgpm$ favours $J^{PC}=1^{++}$ or $2^{-+}$~\cite{bib-Abulencia:2006ma,bib-Belle:2011}.
The inclusive production cross section of the $\Xnew$ resonance has been measured by the LHCb experiment~\cite{bib-lhcb-X3872}.
At the Tevatron it was observed that the $\Xnew$ is produced both through ``prompt'' processes, in which the $\Xnew$ resonance
is created directly, and through decays of B hadrons~\cite{bib-cdf-longlived}, generally referred to as ``nonprompt''. Experimentally, nonprompt processes are distinguishable through the displacement of the $\Xnew$ decay vertex from the primary vertex.
Prompt production of quarkonium states in proton-proton collisions is usually described in the framework of nonrelativistic quantum chromodynamics (NRQCD)~\cite{bib-Brambilla:2010cs}. Quantitative predictions have been calculated for the differential production cross section of the $\Xnew$ in $\Pp\Pap$ collisions at the Tevatron and pp collisions at the Large Hadron Collider (LHC)~\cite{bib-Polosa:2009,bib-Artoisenet:2009wk}. Measurement of the prompt production rate at the LHC as a function of transverse momentum provides a test of the NRQCD factorization approach to $\Xnew$ production.

In this paper a measurement of the differential $\Xnew$ production cross section is presented using decays into $\JPsi \Pgpp\Pgpm$, with the subsequent decay of the \JPsi into a pair of muons.
The analysis makes use of pp collision data recorded by the Compact Muon Solenoid (CMS) experiment at the LHC in 2011, at a centre-of-mass energy of 7\TeV, corresponding to an integrated luminosity of 4.8\fbinv.
The analysis is performed in the kinematic range of \pt of the $\JPsi \Pgpp\Pgpm$ system between 10 and 50\GeV and the rapidity within $|y|<1.2$. The cross section measurement proceeds by determining
the ratio of the $\Xnew$ and $\psitwo$ cross sections, where both states decay to $\JPsi \Pgpp\Pgpm$.
In this ratio, systematic uncertainties common to both states largely cancel, either partially, as those related to the trigger and the reconstruction of the \JPsi mesons, or fully, as for the integrated luminosity.
The fraction of nonpromptly produced $\Xnew$ states is also measured. The cross section times branching fraction for prompt $\Xnew$ production with $\JPsi \Pgpp\Pgpm$ in the final state is then extracted by using a previous CMS measurement of the differential cross section for prompt $\psitwo$ production in the same kinematic range~\cite{bib-jpsi}.
The differential cross section for prompt $\Xnew$ production times the branching fraction is determined for the first time  as a function of transverse momentum.
Finally, the invariant-mass distribution of the dipion system in $\Xnew \rightarrow \JPsi \Pgpp\Pgpm$ decays is studied.

Throughout the analysis, the corrections for detector acceptances and efficiencies are determined under the assumption that the $\Xnew$ has quantum numbers $J^{PC}= 1^{++}$ and that both the $\Xnew$ and the $\psitwo$ are unpolarized. The unknown polarizations of the $\Xnew$ and the $\psitwo$ lead to large uncertainties, in particular in the acceptance of the final-state muon pair for extreme polarization hypotheses.

This paper is structured as follows: after a brief description of the CMS detector in Section~\ref{sec:cms}, the data sample and event selection are discussed in Section~\ref{sec:sel}. The measurement of the cross section ratio is reported in Section~\ref{sec:inclusive}. Section~\ref{sec:nonprompt} gives the measurement of the relative fraction of nonprompt $\Xnew$ production. In Section~\ref{sec:promptxsec} the cross section for prompt $\Xnew$ production is presented. Finally, in Section~\ref{sec:dipionmass}, the $\Pgpp\Pgpm$ mass spectrum in $\Xnew$ to $\JPsi \Pgpp\Pgpm$ decays is reported.

\section{CMS detector}\label{sec:cms}

The central feature of the CMS apparatus is a superconducting solenoid of 6\unit{m} internal diameter. Within the solenoid, in a 3.8\unit{T} magnetic field, are a silicon pixel and strip tracker, a lead tungstate crystal electromagnetic calorimeter, and a brass/scintillator hadron calorimeter. 
The main subdetectors used in this analysis are the silicon tracker and the muon system.
Charged-particle trajectories are measured in the pseudorapidity region $|\eta|<2.5$, where $\eta = -\ln[\tan(\theta/2)]$, with $\theta$ being the polar angle with respect to the anticlockwise-beam direction. The tracker provides an impact parameter resolution of $\approx$15\mum.
Muons are detected in three types of gas-ionization detectors embedded in the steel flux return yoke: drift tubes in the barrel, cathode strip chambers in the endcaps, and resistive plate chambers in both the barrel and endcaps.
Matching the muons to the tracks measured in the silicon tracker results in a transverse momentum resolution between 1 and 1.5\%, for \pt values up to 50\GeV.
A two-level trigger system selects relevant pp collision events for offline reconstruction. The first level (L1) of the CMS trigger system is composed of custom hardware processors. The L1 trigger conditions are adjusted such as to limit the trigger rate to less than 100~kHz. The high-level trigger (HLT) runs on a processor farm to further reduce the rate to a few 100~Hz before data storage.
A detailed description of the detector can be found elsewhere~\cite{JINST}.

\section{Event selection}\label{sec:sel}

The event selection criteria are largely driven by requirements imposed at the trigger level.
At both trigger levels, at least two muons are required. At the HLT, events are accepted if the two muons are of opposite charge, have an invariant mass between 2.95 and 3.25\GeV, vertex fit $\chi^2$ probability greater than 0.5\%, and rapidity $|y(\mu^+\mu^-)| < 1.25$. In 2011, the transverse momentum threshold at the trigger level for the dimuon system was initially 6.9\GeV, which was increased to 9.9\GeV near the end of data taking.  In addition, to cope with increasing
instantaneous luminosities, events in which two muons bend toward each
other in the magnetic field were rejected by criteria added near the
beginning of the data taking.
 The data sample consists of events where an average of six pp collisions in the same bunch crossing (pileup) occur.

In the offline event selection, similar criteria are imposed on the muon pair. Muons are required to have opposite sign. The rapidity of the muon pair is required to be $|y(\mu^+\mu^-)| < 1.25$.
For the first part of the data, 2011a, corresponding to an integrated luminosity of 2.1\fbinv, a minimum dimuon transverse momenta of $\pt(\mu^+\mu^-) = 7\GeV$ is required.
For the second part of the data, 2011b, corresponding to an integrated luminosity of 2.7\fbinv, the transverse momentum threshold is increased to $\pt(\mu^+\mu^-) = 10\GeV$.
The muon identification criteria are very similar to those used in a previous CMS analysis~\cite{bib-jpsi}.
Each candidate muon track must be matched to a triggered muon and have a transverse momentum $\pt(\mu) >4$\GeV in the central-pseudorapidity interval $|\eta(\mu)|<1.2$, or $\pt(\mu) >3.3$\GeV in the forward region $1.2<|\eta(\mu)|<2.4$. These requirements, together with the rapidity and transverse momentum selection criteria on the muon pair, define the \JPsi acceptance, $A(\JPsi)$.
Each muon track must have at least 11 tracker hits, of which at least two are in the silicon pixel layers. The tracks are required to intersect the beam line within a cylinder of 3\unit{cm} in radius and 30\unit{cm} in length around the primary vertex position, selected as the vertex with the largest sum of $\pt^2$ of the tracks associated with it.
The track fit is required to have a $\chi^2$ per degree of freedom, $\chi^2/\mathrm{ndf}$, smaller than 1.8.
The dimuon vertex fit probability is required to be above $1\%$. The invariant mass of the muon pair is required to be within 75\MeV of the fitted $\JPsi$ peak, corresponding to  about ${\pm}2.5$ times the detector resolution
 for the \JPsi mass region. Based on this selection, an almost background-free sample of about ten million reconstructed \JPsi candidates is obtained.

The $\JPsi \Pgpp \Pgpm$ system is reconstructed by combining the candidate muon tracks from each candidate \JPsi with pairs of oppositely charged tracks, which are assumed to be pions. 
 Each $\mu^+ \mu^- \Pgpp \Pgpm$ combination is refitted, constraining the four tracks to come from a common vertex and the muon-pair invariant mass to the \JPsi mass~\cite{bib-pdg}. Combinations yielding a vertex fit probability smaller than 5\% are rejected to suppress combinatorial background.
The pion tracks must have a fit $\chi^2/\mathrm{ndf} < 5$ and contain at least two (seven) silicon pixel (strip) hits. The refitted pion tracks must also have a transverse momentum larger than 600\MeV.

Random combinations of tracks form a significant combinatorial background to the $\Xnew$ and $\psitwo$ signals. For signal events, the pions are expected to have a direction close to that of the \JPsi candidate. Exploiting this property, the combinatorial background is reduced by requiring the distance $\Delta R = \sqrt{(\Delta \eta)^2 + (\Delta \phi)^2}$, where $\Delta \eta$ and $\Delta \phi$ are the pseudorapidity and azimuthal angle differences between the pion and the \JPsi candidate momenta, to be smaller than 0.55.
The requirement $\Delta R<0.55$, together with the pion transverse momentum selection, define the dipion acceptance, $A(\Pgpp\Pgpm)$.
The event selection criteria are driven by studies from simulation, whose description is reported below, using
the quantity $\mathrm{S}/\sqrt{\mathrm{S}+\mathrm{B}}$, where S and B represent the numbers of signal and background candidates, respectively, in a $\pm 2\sigma$ window around the $\Xnew$ mass, and $\sigma$ is the mass resolution (about 6\MeV). 
 In addition, the Q value of the decay is required to be smaller than 300\MeV, where $\mathrm{Q} = m(\mu^+\mu^-\Pgpp\Pgpm) - m(\JPsi)^{\mathrm{PDG}} - m(\Pgpp\Pgpm)$, with $m(\mu^+\mu^-\Pgpp\Pgpm)$ being the invariant mass of the $\JPsi \Pgpp \Pgpm$ system, $m(\Pgpp\Pgpm)$ the invariant mass of the pion pair, and $m(\JPsi)^{\mathrm{PDG}}$ the world-average \JPsi mass~\cite{bib-pdg}.
This selection criterion constrains the mass of pion pairs from $\Xnew$ decays to values larger than about 470\MeV and removes about 20\% of the remaining background, while retaining 97\% of the $\Xnew$ signal, as determined from simulation. 
The invariant mass of the pion pair measured in data, as shown in Section~\ref{sec:dipionmass}, has a negligible contribution below 500\MeV , and thus no bias is introduced with the Q-value requirement.

The $\JPsi \Pgpp \Pgpm$ candidate is required to be in the rapidity region $|y|<1.2$, to have transverse momentum $\pt < 50$\GeV and
 $\pt > 10$\GeV for the period 2011a or $\pt > 13.5$\GeV for the period 2011b.
The resulting data sample consists of about 1.9 million $\JPsi \Pgpp \Pgpm$ candidates with invariant mass between 3.6 and 4.0\GeV. 
The average number of $\JPsi \Pgpp \Pgpm$ candidates per event for events with at least one such candidate 
 is reduced from 7.8 to 2.2 after the event selection.

\begin{figure*}[bhtp]
  \begin{center}
   \includegraphics[width=0.6\textwidth]{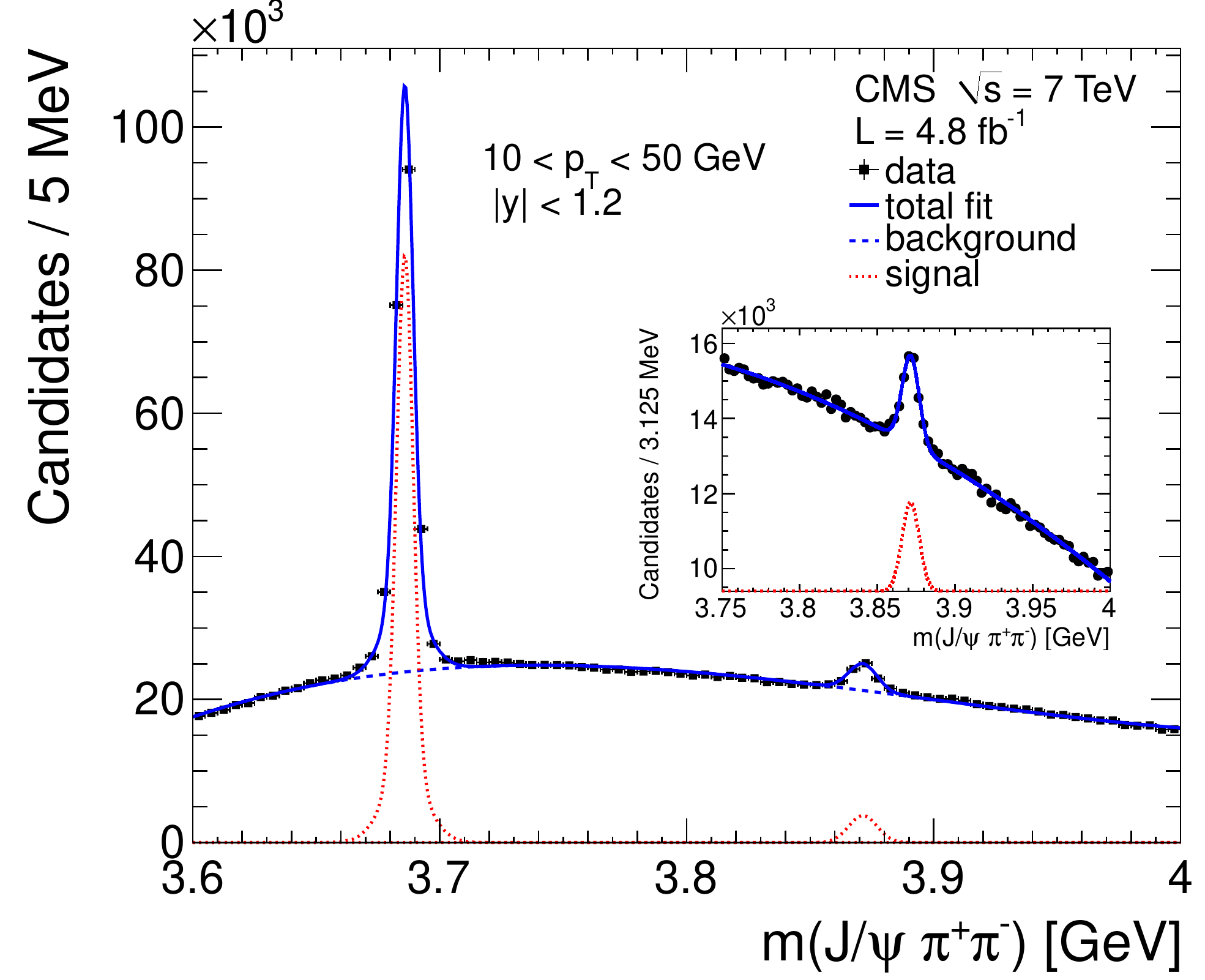}
   \caption{The $\JPsi \Pgpp\Pgpm$ invariant-mass spectrum for $10 < \pt < 50$\GeV and $|y|<1.2$. The lines represent the signal-plus-background fits (solid), the background-only (dashed), and the signal-only (dotted) components. The inset shows an enlargement of the $\Xnew$ mass region.}
    \label{fig:mBoth}
  \end{center}
\end{figure*}

 Figure~\ref{fig:mBoth} shows the invariant-mass distribution of the $\JPsi \Pgpp\Pgpm$ candidates passing the full event selection. An unbinned maximum-log-likelihood fit is used, where Gaussian distributions describe the signals and a Chebyshev polynomial the background. More details about the fit will follow in Section~\ref{sec:inclusive}. Clear $\psitwo$ and $\Xnew$ signals are observed with widths of about 5\MeV and 6\MeV, respectively, dominated by the detector resolution and consistent with simulation.

Detailed event simulations are used to determine detector effects such as acceptances, efficiencies, and resolutions. Events containing $\Xnew$ or $\psitwo$ states are generated using \PYTHIA~\cite{bib-pythia}, and decayed using \EVTGEN~\cite{bib-evtgen}, with the signal resonances forced to decay into the $\JPsi \Pgpp\Pgpm$ final state. Photon final-state radiation (FSR) is implemented using \PHOTOS~\cite{bib-photos1,bib-photos2}. The $\Xnew$ and $\psitwo$ resonances are assumed to be unpolarized.
Since \PYTHIA does not include the simulation of $\Xnew$ production and decay, the program is modified to use the $\chi_{c1}$ particle with its mass set to 3871.6\MeV.
The $\chi_{c1}$ particle has the quantum numbers $J^{PC}=1^{++}$, corresponding to those favoured for the $\Xnew$~\cite{bib-XJPC,bib-Brambilla:2010cs}.
Simulated events for prompt production are used as the baseline. Events with B-hadron decays are simulated and used in the $\Xnew$ nonprompt-fraction measurement.
The $\Xnew \rightarrow \JPsi \Pgpp \Pgpm$ decay is generated with an intermediate $\rho^{0}$ resonance, as suggested by previous measurements~\cite{bib-CDFangluar:2005,bib-Belle:2011} and confirmed in this analysis (Section~\ref{sec:dipionmass}).
In \EVTGEN a two-body phase-space decay is used for the $\Xnew \rightarrow \JPsi \rho^{0}$ decay,
and the $\rho^{0}$ decay to a pair of pions is generated with decay-angle distributions reflecting their respective spins.
A nonresonant $\Xnew \rightarrow \JPsi \Pgpp \Pgpm$ decay is also considered using the \EVTGEN model for the $\psitwo \rightarrow \JPsi \Pgpp \Pgpm$ decay. 
The study of systematic uncertainties uses a version of \PYTHIA that includes colour-octet contributions with NRQCD matrix elements, as determined from CDF data~\cite{bib-octect1,bib-octect2}.

 Large samples of simulated events are produced separately for the $\Xnew$ and $\psitwo$ resonances, both for prompt production and nonprompt production in B-hadron decays. 
The response of the detector is simulated in detail using \GEANTfour~\cite{GEANT4}. The simulated samples are processed through the trigger emulation and event reconstruction of the CMS experiment, without taking into account other pp collisions in the same bunch crossing (pileup) since the analysis is not sensitive to it, as discussed in Section~\ref{sec:inclusive}.

\section{Measurement of the cross section ratio\label{sec:inclusive}}

The ratio of the cross section times the $\JPsi \Pgpp\Pgpm$ branching fraction is obtained from the measured numbers of signal events for $\Xnew$ and $\psitwo$, $N_{\Xnew}$ and $N_{\psitwo}$, correcting for the efficiency ($\epsilon$) and acceptance ($A$) estimated from simulations, according to

\begin{equation}
\label{eq:R}
R=\frac{\sigma(\Pp\Pp \rightarrow \Xnew + \text{anything}) \cdot \mathcal{B}(\Xnew \rightarrow \JPsi \Pgpp\Pgpm)}
     {\sigma(\Pp\Pp \rightarrow \psitwo + \text{anything}) \cdot \mathcal{B}(\psitwo \rightarrow \JPsi \Pgpp\Pgpm)}
 =\frac{N_{\Xnew}\cdot A_{\psitwo} \cdot \epsilon_{\psitwo}}{N_{\psitwo}\cdot A_{\Xnew} \cdot\epsilon_{\Xnew}}.
\end{equation}

The acceptance corrections account for the kinematic reach of the dimuon trigger and the angular acceptance of the CMS detector.
These corrections depend on assumptions about the angular distribution of the final-state muon and pion pairs.
To minimize the effect of these assumptions, the measurement is also presented as a ``fiducial'' cross section ratio, defined as

\begin{equation}
\label{eq:Rfid}
R_\text{fiducial}=\frac{N_{\Xnew}\cdot \epsilon_{\psitwo}}{N_{\psitwo}\cdot \epsilon_{\Xnew}} ,
\end{equation}

within a phase-space window with the following kinematic requirements on the muons, dimuons, and pions: muons with
$\pt(\mu)> 4$\GeV for $ |\eta(\mu)|<1.2$ and $\pt(\mu) > 3.3$\GeV for $1.2<|\eta(\mu)|<2.4$;
$\pt(\mu^+\mu^-) >7\GeV$ and $|y(\mu^+\mu^-)| < 1.25$ for the dimuons;
each pion with transverse momentum greater than 600\MeV and a distance with respect to the dimuon $\Delta R < 0.55$.

The signal yields are determined from unbinned maximum-likelihood fits to the invariant-mass spectra of the $\JPsi \Pgpp\Pgpm$ system, separately for the $\Xnew$ and $\psitwo$, in the mass windows 3.75--4\GeV and 3.6--3.8\GeV, respectively, and in five bins of $\pt$ with edges: 10, 13.5, 15, 18, 30, and 50\GeV. Following the evolution of the trigger thresholds with time, the first bin in transverse momentum, 10--13.5\GeV, includes only data from the period 2011a, while for \pt bins above 13.5\GeV, the full dataset (2011a+2011b) is used.
The inclusive signal yield for \pt between 10 and 50\GeV is determined by combining the first \pt bin from 2011a, weighted to account for luminosity and trigger differences, with the remaining bins from the full dataset.

\begin{figure*}[tp]
\begin{center}
	\includegraphics[width=0.48\textwidth]{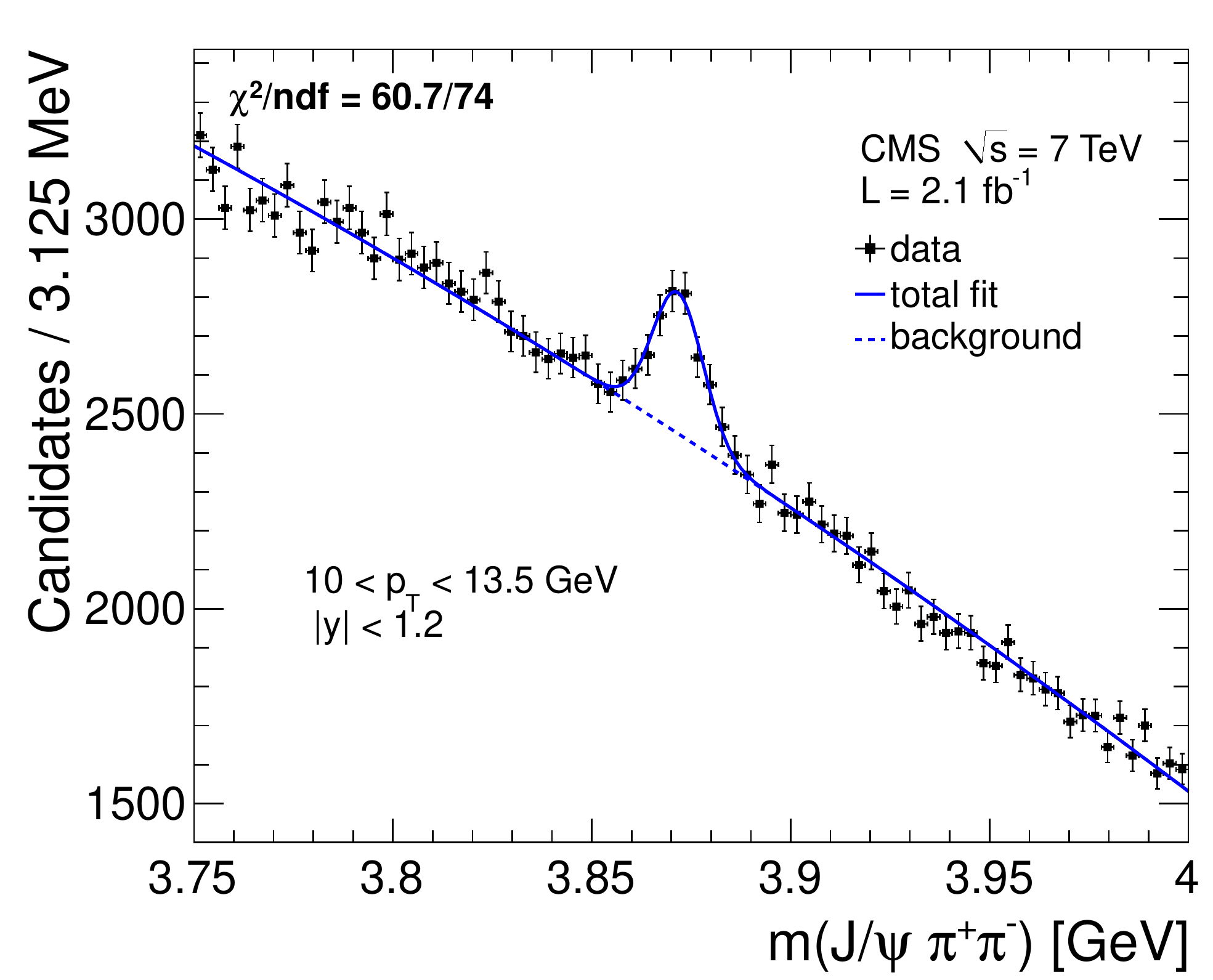}
	\includegraphics[width=0.48\textwidth]{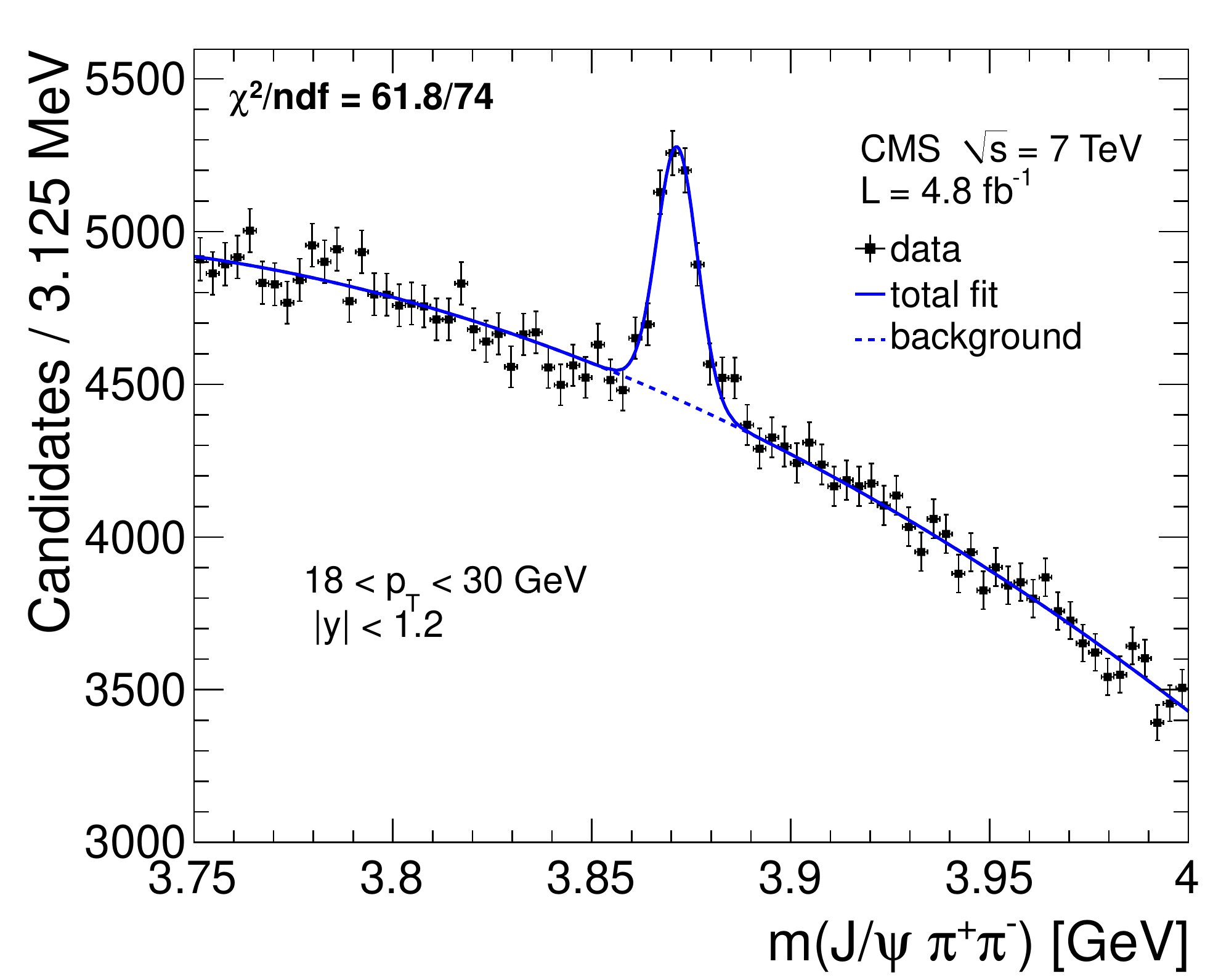}
\caption{ The $\JPsi \Pgpp\Pgpm$ invariant-mass distribution in the $\Xnew$ region for two bins of transverse momentum, 10--13.5\GeV (left) and 18--30\GeV (right). The lines represent the signal-plus-background fits (solid) and the background-only components (dashed). The $\chi^2/\mathrm{ndf}$ of the fit is also reported. 
}
\label{fig:ptexamples}
\end{center}
\end{figure*}

\begin{table}[b]
\centering
\topcaption[]{\label{tab:YieldAccEff} Measured numbers of signal events, $N_{\Xnew}$ and $N_{\psitwo}$, and the ratios of the $\Xnew$ and $\psitwo$ efficiencies ($\epsilon$) and acceptances ($A$) as a function of the $ \JPsi \Pgpp\Pgpm$ \pt. For the first transverse momentum bin only the data from period 2011a are included. All uncertainties are statistical only.
}
{\small
\begin{tabular}{|l|c|r|c|c|c|}
\hline
Dataset & \pt (\GeVns{})  & $N_{\Xnew}$ &  $N_{\psitwo}$ &  $\frac {\epsilon_{\psitwo}}{\epsilon_{\Xnew} }$ & $\frac {A_{\psitwo} \cdot \epsilon_{\psitwo}}{A_{\Xnew} \cdot \epsilon_{\Xnew} }$ \\
\hline
2011a   & 10--13.5 & 1850 $\pm$ 200 &  25 450 $\pm$ 330  & 1.055 $\pm$ 0.011 & 0.999 $\pm$ 0.025  \\
2011a+b & 13.5--15 & 1700 $\pm$ 170 & 24 130 $\pm$ 440 & 1.032 $\pm$ 0.014 & 0.951 $\pm$ 0.025 \\
2011a+b & 15--18 & 2770 $\pm$ 210  & 39 450 $\pm$ 470 & 1.031 $\pm$ 0.011 & 0.979 $\pm$ 0.020  \\
2011a+b & 18--30 & 3360 $\pm$ 230 & 56 920 $\pm$ 510  & 1.035 $\pm$ 0.011 & 1.019 $\pm$ 0.018    \\
2011a+b & 30--50 & 860 $\pm$ 140 & 12 130 $\pm$ 230 & 1.052 $\pm$ 0.037 & 1.103 $\pm$ 0.056      \\
\hline
2011a+b & 10--50 & 11 910 $\pm$ 490 & 178 540 $\pm$ 850 & 1.040 $\pm$ 0.006 &  0.984 $\pm$ 0.017  \\
\hline
\end{tabular}
}
\end{table}
 In the fits, the $\psitwo$ resonance shape is parametrized using two Gaussian functions with a common mean, while a single Gaussian is used for the $\Xnew$ signal. The nonresonant background is fitted with a second-order Chebyshev polynomial. The free parameters in the fit are the signal and background yields, the mass and widths of the Gaussian functions, the fraction of signal associated with each Gaussian, and two background-shape parameters.
Figure~\ref{fig:ptexamples} shows examples of fitted mass distributions for a low- and a high-transverse-momentum bin.
The measured numbers of $\Xnew$ and $\psitwo$ signal events are listed in Table~\ref{tab:YieldAccEff}. 

The acceptances and efficiencies of the $\Xnew$ and $\psitwo$ final states are factorized into four components, each of which is determined individually from the simulation: the acceptance $A(\JPsi)$ and efficiency $\epsilon(\JPsi)$ for the trigger and detection of the \JPsi, and the acceptance $A(\Pgpp\Pgpm)$ and efficiency $\epsilon(\Pgpp\Pgpm)$ for the pion pair, including the $\JPsi \Pgpp\Pgpm$ vertex probability requirement.
The acceptances are the same for the 2011a and 2011b datasets for the \pt bins in common ($\pt > 13.5$\GeV).
The efficiency is calculated for the 2011a dataset in each bin since the changes in efficiency related to the trigger evolution during data taking do not affect the efficiency ratio.
The average value of $A \cdot \epsilon$ in each \pt bin is determined using fine-grained bins in transverse momentum as

\begin{equation}
\label{eqn:AccEffCorr}
  \left< \frac {1}{A \cdot \epsilon } \right>_\text{bin} \equiv \left. \sum\limits_{i=1}^{N^\text{bin}_\text{fine}} \frac{N_i}{A^i \cdot \epsilon^i} \middle/ \sum\limits_{i=1}^{N^\text{bin}_\text{fine}} N_i,  \right. \\
\end{equation}

where $N_i$ is the number of signal events observed in the data, $A^i=A^i(\JPsi)\cdot A^i(\Pgpp\Pgpm)$, $\epsilon^i=\epsilon^i(\JPsi)\cdot\epsilon^i(\Pgpp\Pgpm)$ are the acceptance and efficiency in each fine bin, and $N^\text{bin}_\text{fine}$ is the number of fine bins contained in each \pt interval.  This procedure accounts for the large variation in acceptance and efficiency over the wide \pt bins, relying on the \pt spectrum from the data. The number of signal events in each fine bin is determined using a sideband-subtraction technique. 
 The ratios of the acceptances and efficiencies, listed in Table~\ref{tab:YieldAccEff}, are different from unity because of small differences in the $\Xnew$ and $\psitwo$ decay kinematics.

Studies are performed to verify the description of the data by the simulations and to determine the systematic uncertainties. These are listed in Table~\ref{tab:sys} and described in the following. 

\begin{itemize}
\item \textit{Fit functions}.
The systematic uncertainty in the signal extraction from the invariant-mass spectrum is determined by variation of the fit parametrization independently for the $\Xnew$ and $\psitwo$. Using a third-order Chebyshev polynomials for the backgrounds or the sum of a Gaussian and a Crystal Ball~\cite{bib-Oreglia} function for the signal, variations of 1--2\% 
are found. Fixing the $\Xnew$ and $\psitwo$ mass difference to the PDG value~\cite{bib-pdg} in the fit changes the result by less than 1\%.
\item \textit{Muon-pair efficiency}.
Systematic uncertainties in muon efficiencies largely cancel in the cross section ratio measurement.
Single-muon efficiencies are determined from \JPsi events using a tag-and-probe technique on both the data and simulation~\cite{bib-jpsi}.
The systematic uncertainty in the cross section ratio from this source is less than 1\%.
\item \textit{Pion-pair efficiency}.
The systematic uncertainty in the efficiency for the reconstruction of the pion pair is determined by comparison of the measured and simulated event yields from $\psitwo \rightarrow \JPsi \Pgpp\Pgpm$ and $\psitwo \rightarrow \mu^+\mu^-$ decays. After corrections for the branching fractions~\cite{bib-pdg} and differences in the acceptance and efficiency for the muon pair, the ratio of event yields in the two decay channels differs from unity because of different dipion reconstruction efficiencies. The more precisely measured value $\mathcal{B}(\psitwo\rightarrow e^+ e^-)$~\cite{bib-pdg} is used, instead of that for $\psitwo\rightarrow \mu^+ \mu^-$, assuming lepton universality.
 Comparison of the simulation with the data reveals differences in dipion efficiency of 5\% for $\pt < 15$\GeV and 1\% at higher transverse momentum.
\begin{table}[bt]
\centering
\topcaption[]{\label{tab:sys} Summary of the relative systematic uncertainties for $R_\text{fiducial}$ and $R$. The variation over the \pt bins is given. The systematic sources common to both $R_\text{fiducial}$ and $R$ are reported at the top, followed by those affecting only $R_\text{fiducial}$ and only $R$.
}
{\small
\begin{tabular}{|l|c|}
\hline
Source & Relative uncertainty (\%) \\
\hline\hline
 Common to $R_\text{fiducial}$ and $R$ & \\
\hline
Fit functions & 1--2 \\
$\epsilon(\mu^+\mu^-)$ & $<$ 1\\
$\epsilon(\Pgpp\Pgpm)$ & 1--5 \\
Efficiency statistical precision & 1--3 \\
\hline\hline
 Specific to $R_\text{fiducial}$ &  \\
\hline
$\Xnew$  \pt spectrum & 2--5 \\
$\psitwo$ \pt spectrum & 1--4 \\
\hline
Total systematic uncertainty in $R_\text{fiducial}$ & 4--8 \\
\hline\hline
 Specific to $R$ & \\
\hline
$\Xnew$  \pt spectrum & 1--11 \\
$\psitwo$ \pt spectrum & 1--4 \\
$m(\Pgpp\Pgpm)$ spectrum & 1--2 \\
Acceptance statistical precision & 1--3\\
\hline
Total systematic uncertainty in $R$ & 5--13 \\
\hline
\end{tabular}
}
\end{table}
\item \textit{Efficiency statistical precision}.
The efficiency uncertainties introduced by the statistical limitations of the simulated samples is less than 1\% in general, rising to 3\% for $30 < \pt < 50$\GeV.
\item \textit{$\Xnew$ \pt spectrum}.
The dependence of the measurement on the transverse momentum spectrum of the $\Xnew$ is estimated by repeating  the analysis with a simulation including colour-octet contributions~\cite{bib-octect1,bib-octect2}. Simulations with and without colour-octet contributions lead to large variations of the \pt spectra that are still compatible with the data. 
The differences between these two cases, 2--5\% on $R_\text{fiducial}$ and 5--6\% on $R$, are taken as the systematic uncertainty.
Variations of similar size are obtained when reweighting the simulated $\Xnew$ \pt spectrum to match the data.
The uncertainty in the \pt spectrum extracted from the data is also considered as a source of systematic uncertainty, which is added in quadrature.
The uncertainty in $R_\text{fiducial}$ is found to be 2--5\% . 
The rapidly changing acceptance as a function of transverse momentum makes the $R$ measurement very sensitive to the \pt spectrum, in particular for low transverse momentum and for the \pt-integrated result. The uncertainty in $R$ is 11\% in the first \pt bin and 1--7\% elsewhere.
\item \textit{$\psitwo$ \pt spectrum}.
For the $\psitwo$, the simulated \pt spectrum is reweighted to match the distribution observed in data, and the efficiency and acceptance corrections are recalculated. The change in the cross section ratios, both for $R_\text{fiducial}$ and $R$, is about 4\% in the lowest transverse momentum bin and 1--3\% elsewhere.
\item \textit{$m(\Pgpp\Pgpm)$ spectrum}.
The dipion invariant-mass spectrum of the $\Xnew \rightarrow \JPsi \Pgpp\Pgpm$ decay is extracted from the data, as described in Section~\ref{sec:dipionmass}, and compared with the expectations from the simulated samples. The dependence of the efficiency corrections on the dipion invariant mass is weak, and the systematic uncertainty in $R_\text{fiducial}$ is negligible.
The dependence of $R$ on the assumed invariant-mass spectrum of the pion pair is estimated by reweighting the generated dipion invariant-mass spectrum to match the data. This leads to changes in the cross section ratio $R$ of up to 2\%.
\item \textit{Acceptance statistical precision}.
The uncertainty in the estimate of the dimuon and dipion acceptances owing to the statistical limitations of the simulated samples is 1\%, rising to 3\% at high transverse momentum.
\end{itemize}
The stability over time of the $\JPsi  \rightarrow \mu^+\mu^-$ yield relative to the $\psitwo \rightarrow \JPsi \Pgpp\Pgpm$ yield verifies that the muon and track selections used in the analysis 
are not sensitive to beam conditions or the amount of pileup.

Adding all the systematic uncertainties in quadrature, a total systematic uncertainty of 4--8\% in $R_\text{fiducial}$ and 5--13\% in $R$ is obtained.

\begin{figure*}[t]
\begin{center}
\includegraphics[width=0.49\textwidth]{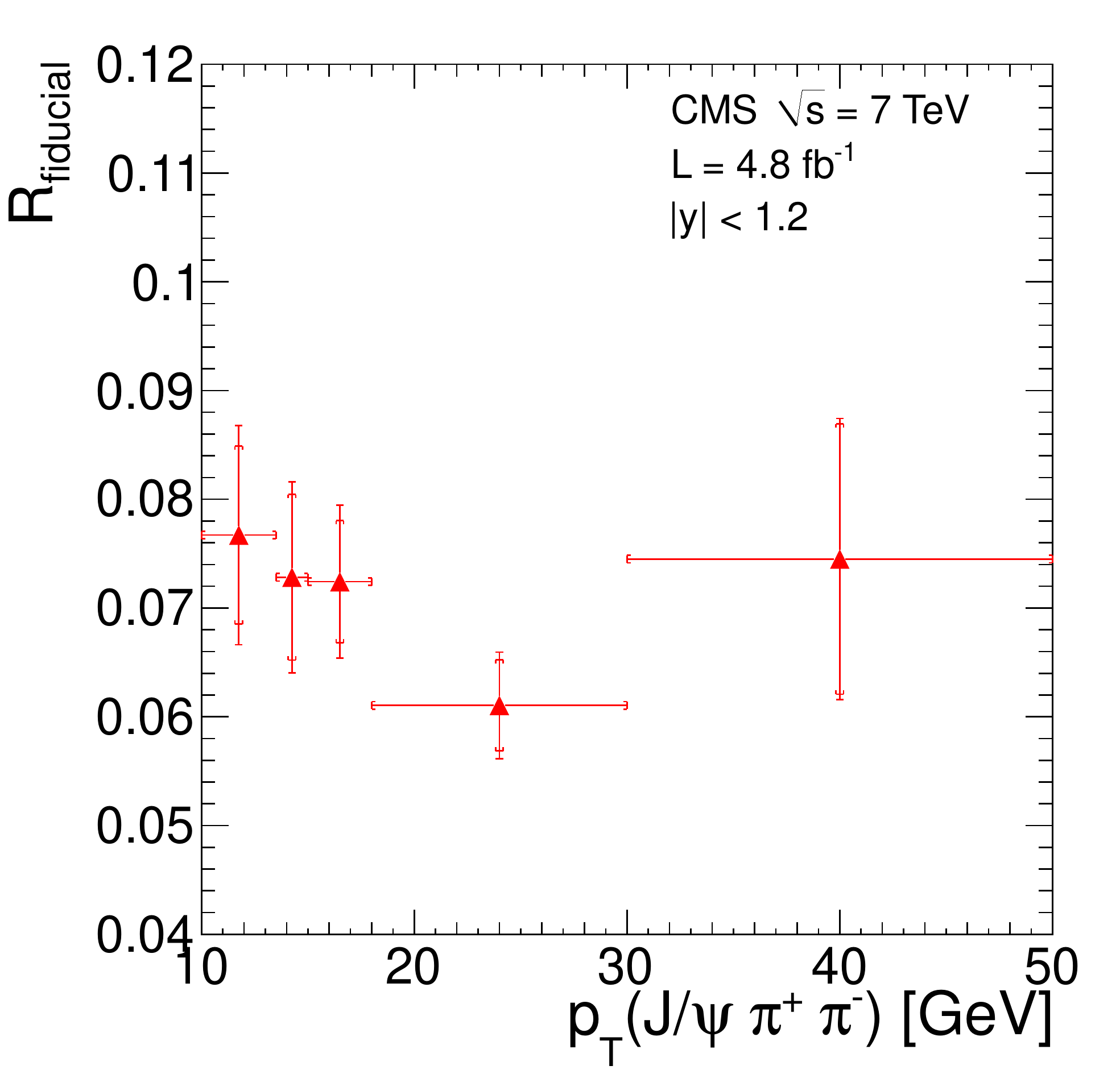}
\includegraphics[width=0.49\textwidth]{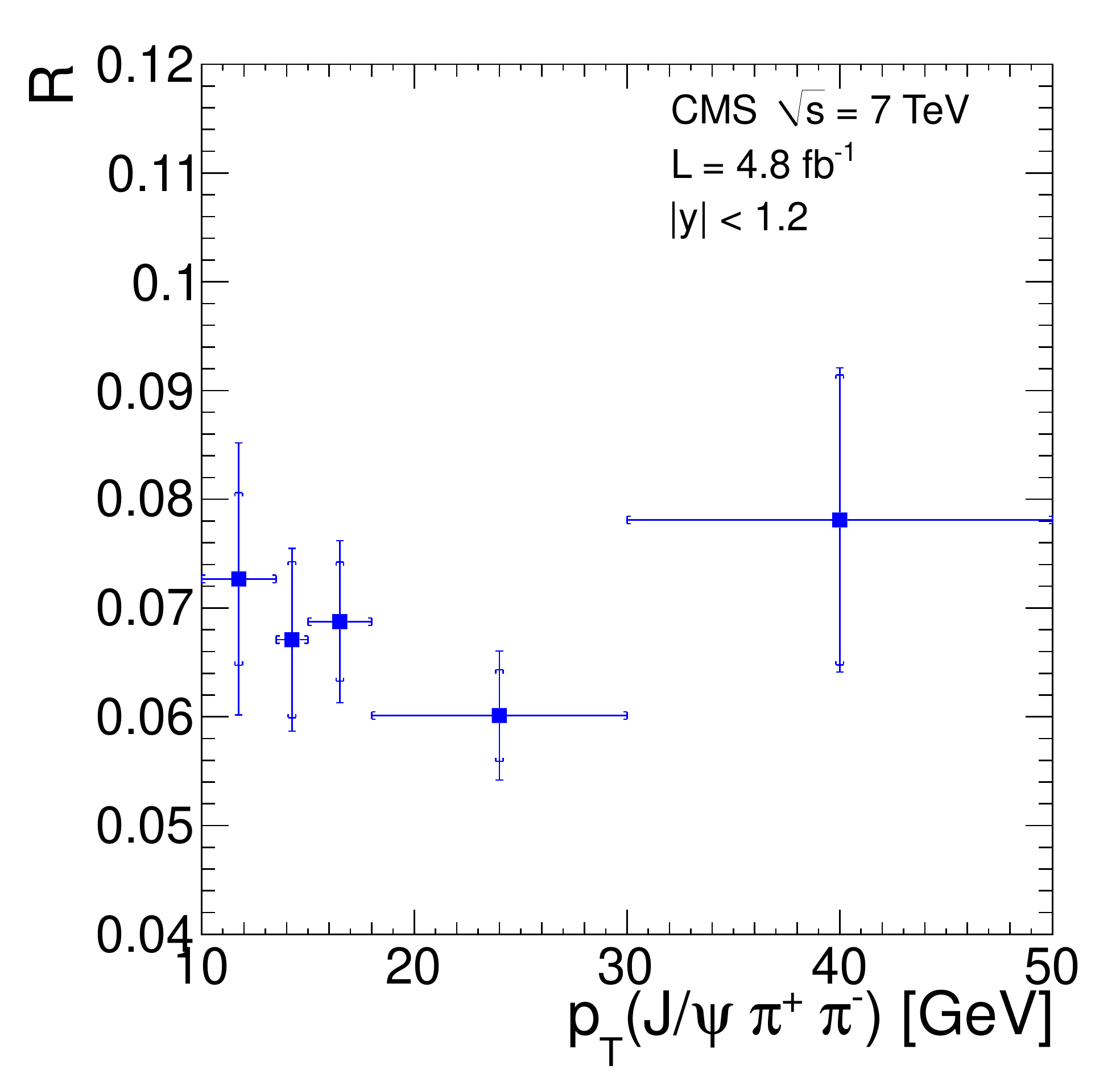}
\caption{Ratios of the $\Xnew$ and $\psitwo$ cross sections times branching fractions, without ($R_\text{fiducial}$, left) and with ($R$, right) acceptance corrections for the muon and pion pairs, as a function of \pt.
The inner error bars indicate the statistical uncertainty and the outer error bars represent the total uncertainty. The data points are placed at the centre of each \pt bin.
}
\label{fig:Rsysvspt}
\end{center}
\end{figure*}

The cross section ratio is determined as a function of the transverse momentum of the $\JPsi \Pgpp\Pgpm$ system.
The results for both $R_\text{fiducial}$ (Eq.~(\ref{eq:Rfid})) and the fully acceptance-corrected $R$ (Eq.~(\ref{eq:R}))  are listed in Table~\ref{tab:R} and shown in Fig.~\ref{fig:Rsysvspt}. No significant dependence on transverse momentum is observed for either quantity. 
These results are obtained under the assumption that the $\Xnew$ quantum numbers are $J^{PC}= 1^{++}$, as favoured by existing data~\cite{bib-XJPC,bib-Brambilla:2010cs}, and no systematic uncertainty is assigned to cover other cases. 

\begin{table}[b]
\centering
\topcaption[]{\label{tab:R}
The ratios of the measured cross sections times branching fractions, $R_\text{fiducial}$ and $R$, as a function of the transverse momentum of the $\JPsi \Pgpp\Pgpm$ system, together with their statistical and systematic uncertainties, respectively.
For the first bin in transverse momentum, only the data from the period 2011a are included.
}
{\small
\begin{tabular}{|l|c|c|c|}
\hline
 Dataset & \pt (GeV) & $R_\text{fiducial}$ & $R$ \\
\hline
 2011a   & 10--13.5  & 0.0767 $\pm$ 0.0082 $\pm$ 0.0059 & 0.0727 $\pm$ 0.0079 $\pm$ 0.0097 \\
 2011a+b & 13.5--15  & 0.0728 $\pm$ 0.0076 $\pm$ 0.0044 & 0.0671 $\pm$ 0.0072 $\pm$ 0.0044 \\
 2011a+b & 15--18  & 0.0724 $\pm$ 0.0056 $\pm$ 0.0042 & 0.0687 $\pm$ 0.0055 $\pm$ 0.0051 \\
 2011a+b & 18--30  & 0.0611 $\pm$ 0.0042 $\pm$ 0.0025 & 0.0601 $\pm$ 0.0042 $\pm$ 0.0042 \\
 2011a+b & 30--50  & 0.075 $\pm$ 0.012 $\pm$ 0.004 & 0.078 $\pm$ 0.013 $\pm$ 0.004 \\
\hline
2011a+b & 10--50  & 0.0694 $\pm$ 0.0029 $\pm$ 0.0036 & 0.0656 $\pm$ 0.0029 $\pm$ 0.0065 \\
\hline
\end{tabular}
}
\end{table}

In the simulations, unpolarized $\Xnew$ and $\psitwo$ states are assumed. 
To evaluate the impact of other polarization scenarios, it is assumed that the $\Xnew$ and the  $\JPsi$ from the $\Xnew \rightarrow \JPsi \Pgpp \Pgpm$ decay have the same polarization.
The polarization of the $\JPsi$ is varied in extreme scenarios, corresponding to fully longitudinal or fully transverse polarization in the helicity and Collins--Soper frames~\cite{bib-Faccioli}. The same variations are performed separately for the $\JPsi$ from $\Xnew$ decays and for the $\psitwo$.
The observed relative shifts of the cross section ratio $R$ are listed in Table~\ref{tab:pol}.
Small effects are found in scenarios where both the $\Xnew$ and the $\psitwo$ have the same polarization.
Assuming only one of the two states is unpolarized leads to variations of about 30\% in the helicity frame and up to 20\% in the Collins--Soper frame.
 Scenarios with transversely polarized $\Xnew$ and longitudinally polarized  $\psitwo$ give variations of up to 90\% for the helicity frame and 30\% for the Collins--Soper frame.
In contrast, the fiducial cross section ratio $R_\text{fiducial}$ is largely insensitive to polarization assumptions, showing maximal variations of 4\%.

\begin{table}[hbt]
\centering
\topcaption[]{\label{tab:pol} Relative variations, in percent, of the
integrated cross section ratio $R$ for different $\Xnew$ and $\psitwo$ polarization hypotheses: transversely (longitudinally) polarized $\JPsi$ are denoted as CST (CSL) in the Collins--Soper frame and HXT (HXL) in the helicity frame. Unpolarized scenarios (labelled unpol) are also included.}
{\small
\begin{tabular}{|l|l|c||l|l|c|}
\hline
\multicolumn{2}{|c|}{Polarization} & Relative & \multicolumn{2}{c|}{Polarization} & Relative \\
 $\Xnew$ &  $\psitwo$  & shifts ($\%$) & $\Xnew$ &  $\psitwo$  & shifts ($\%$) \\
\hline
 CST & CSL & $-$28 & CST & unpol & $-$8  \\
 CSL & CST & $+$31   & CSL & unpol & $+$22 \\
 HXT & HXL & $+$86   & HXT & unpol & $+$28 \\
 HXL & HXT & $-$49 & HXL & unpol & $-$31 \\
\hline
CST & CST & $-$1 & unpol & CST & $+$8 \\
CSL & CSL & $-$5 & unpol & CSL & $-$22  \\
HXT & HXT & $-$6 & unpol & HXT & $-$27 \\
HXL & HXL & $-$1 & unpol &  HXL & $+$25 \\
\hline
\end{tabular}
}
\end{table}

\section{Measurement of the nonprompt fraction}
\label{sec:nonprompt}
\begin{figure*}[tb]
\begin{center}
\includegraphics[width=0.48\textwidth]{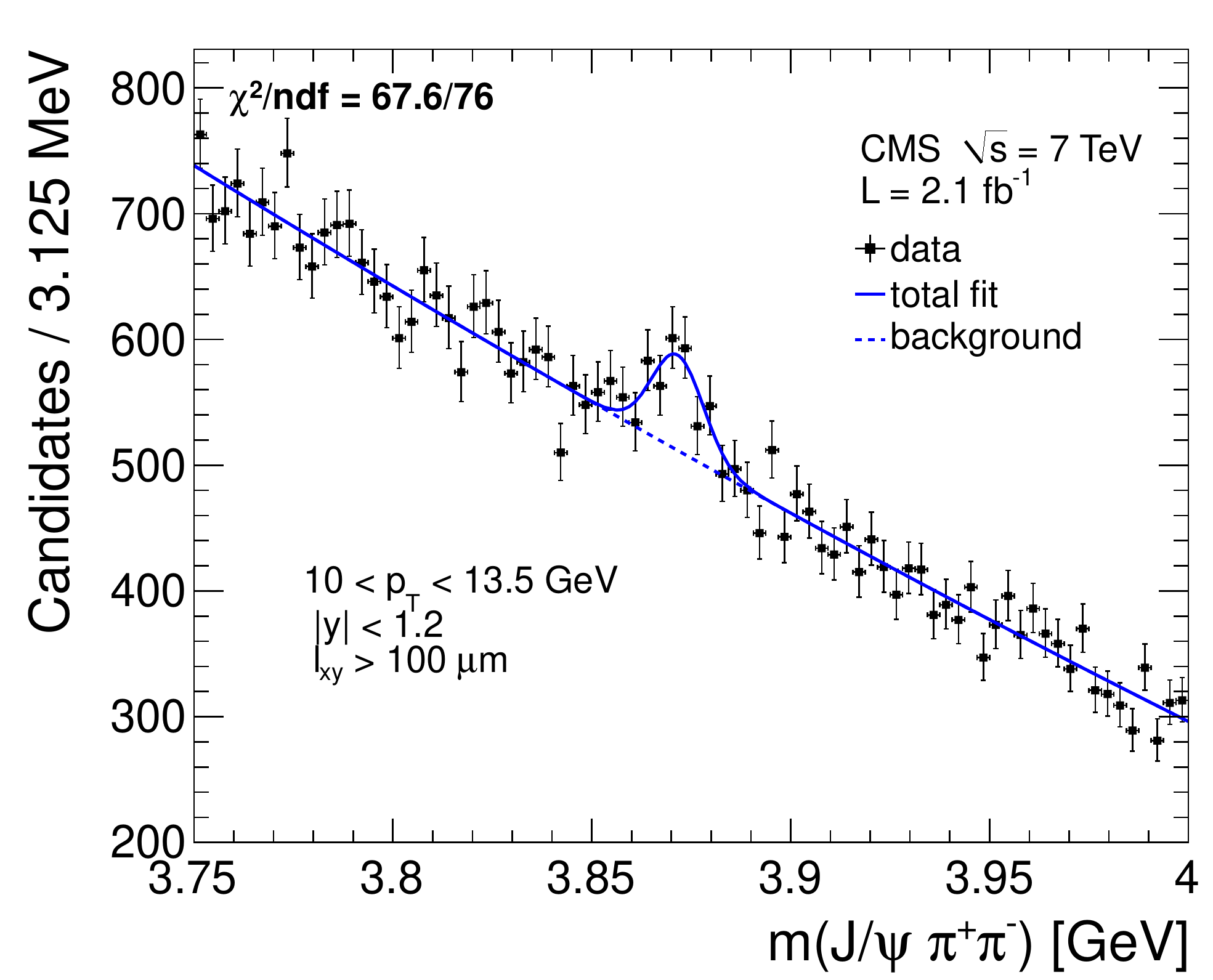}
\includegraphics[width=0.48\textwidth]{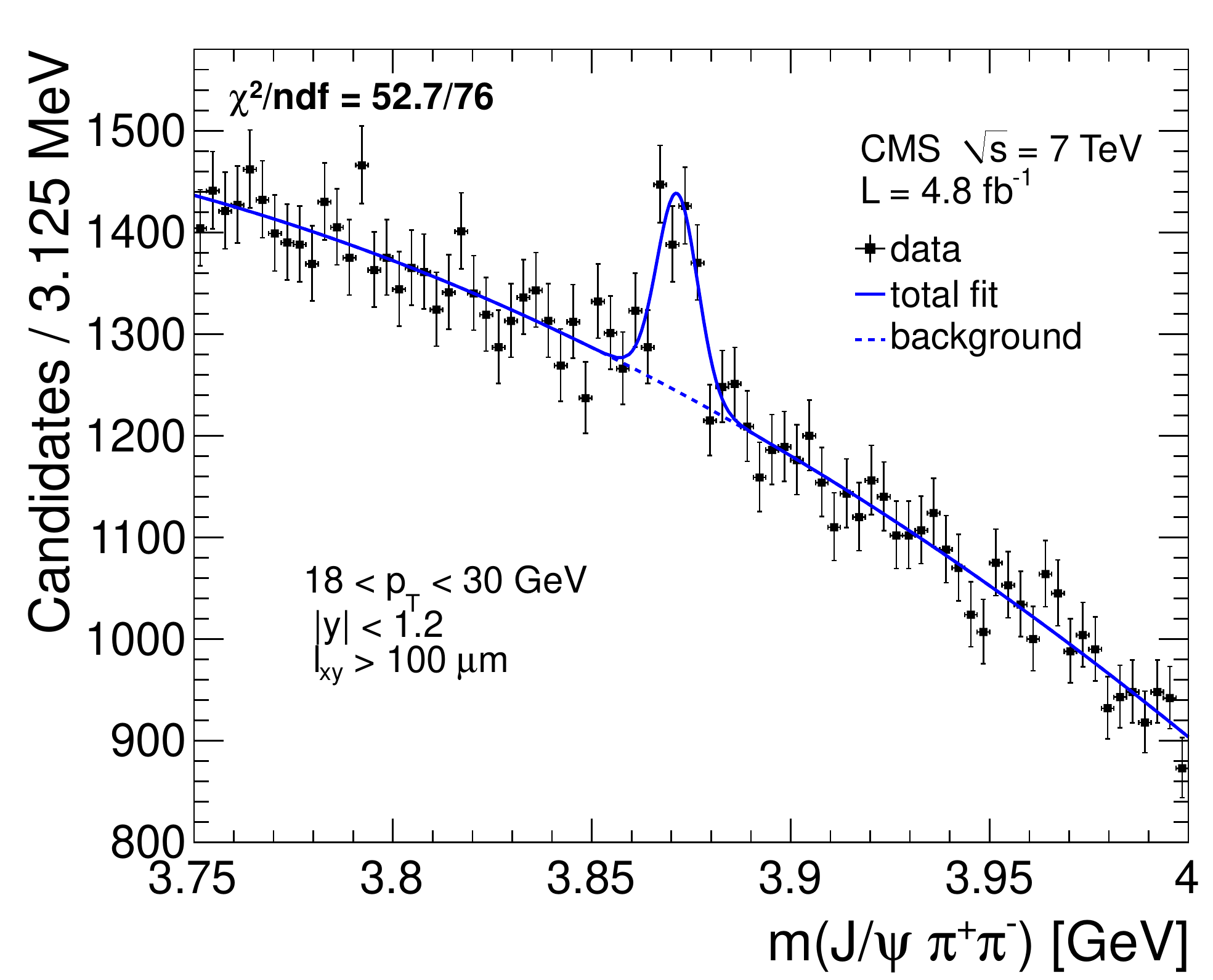}
\caption{The $\JPsi \Pgpp\Pgpm$ invariant-mass distribution in the $\Xnew$ region for \pt $=$ 10--13.5\GeV (left) and 18--30\GeV (right) in a B-enriched sample. The lines represent the signal-plus-background fits (solid) and the background-only components (dashed). The $\chi^2/\mathrm{ndf}$ of the fit is also reported.} 
\label{fig:ptexamples_lxy}
\end{center}
\end{figure*}

The relative contribution to the total $\Xnew$ yield resulting from decays of B hadrons, often referred to as the nonprompt fraction, is determined from the decay lifetime distribution. 
The measurement is performed with the same $\JPsi$ and $\Pgpp\Pgpm$ acceptance criteria presented above.
The ``pseudo-proper'' decay length $\ell_{xy}$ is defined in the plane transverse  to the beam direction as the distance between the vertex formed by the four tracks of the $\JPsi \Pgpp\Pgpm$ system and the closest reconstructed primary vertex along the beam direction, corrected by the transverse Lorentz boost of the $\JPsi \Pgpp\Pgpm$ candidate.
An event sample enriched in $\Xnew$ candidates from B decays is selected by requiring that $\ell_{xy}$ be larger than 100\mum. This selection retains about 80\% of the nonprompt $\Xnew$ candidates, while the contribution from prompt $\Xnew$ is smaller than 0.1\%, as determined from simulation.
The simulated $\ell_{xy}$ distribution is verified using the corresponding distribution from the $\psitwo$ data sample.
The nonprompt fraction is then obtained from the ratio between the signal yields in this B-hadron-enriched sample and the signal yields in the inclusive sample, after correction for the efficiencies of the decay-length-selection criteria, as determined from simulations of prompt and nonprompt $\Xnew$ states. The signal yields are extracted from fits to the $\JPsi \Pgpp\Pgpm$ invariant-mass spectrum, as described in Section~\ref{sec:inclusive}.
In the fits to the B-hadron-enriched sample, the fit parameters for the mass and width are fixed to those determined from the full sample. Figure~\ref{fig:ptexamples_lxy} shows examples of fitted invariant-mass distributions for the B-hadron-enriched sample.

The measurement is found to be stable with respect to variations of the pseudo-proper-decay-length requirement between 50 and 250\mum.
Application of the same method to $\psitwo$ candidates yields the same result as previously measured~\cite{bib-jpsi}.
In an alternative method, similar to the one used in Ref.~\cite{bib-jpsi}, a two-dimensional fit to the invariant mass and the pseudo-proper decay length is performed.
The $\ell_{xy}$ resolution is described by a function that depends on the uncertainty in the pseudo-proper-decay-length measurement, as determined event-by-event from the covariance matrices of the fits to the primary and secondary vertices.
This function is obtained for signal and background, respectively, from the invariant-mass distribution after sideband subtraction~\cite{SBSexample} and from the sideband regions. These distributions are used to fix the lifetime parameters in the two-dimensional fit to correctly model the pseudo-proper-decay-length resolution.
The validity of both methods has also been verified with simulated prompt and nonprompt $\Xnew$ events for the signal, 
 and combining from data the \JPsi and same-sign tracks into a $\JPsi \pi \pi$ candidate for the nonresonant background.
Repeating the analyses on these samples, the nonprompt fractions are consistent with those used in the simulation.
While both methods agree, the method utilizing the requirement on $\ell_{xy}$ is chosen since it has the smaller systematic uncertainty.

\begin{table}[tb]
\centering
\topcaption[]{\label{tab:sysNP} Summary of systematic uncertainties in the $\Xnew$ nonprompt fraction.}
{\small
\begin{tabular}{|l|c|}
\hline
Source & Relative uncertainty (\%)\\
\hline
Vertex estimation   & 1 \\
Background parametrization & 2--3 \\
Efficiency & 3--8 \\
Decay length resolution  & 4 \\
Pileup & 2 \\
\hline
Total systematic uncertainty &  6--10 \\
\hline
\end{tabular}
}
\end{table}

Detailed studies are performed to determine the systematic uncertainties listed in Table~\ref{tab:sysNP} and described in the following.
The selection of the primary vertex is modified by choosing the vertex with the smallest
impact parameter along the 
 beam direction for the $\Xnew$ candidate, instead of the one closest 
to the four-track vertex along the beam direction. This variation changes the measured nonprompt fraction by 1\%.
The systematic uncertainties related to signal extraction, determined by changing the background functions, are 2--3\%.
The difference between the reconstruction efficiency for prompt and nonprompt production, 8\% for the highest transverse momentum bin and
 3--4\% elsewhere, is taken as the systematic uncertainty.
The uncertainty from the simulation of the pseudo-proper-decay-length resolution is estimated by comparing the $\ell_{xy}$ distribution from a simulated $\psitwo$ sample with that from data. The change in the nonprompt fraction when relying on the $\ell_{xy}$ resolution from data is 4\%.
Finally, the systematic uncertainty from the description of pileup events is evaluated from the dependence of the result on the  number of primary vertices in the event and estimated to be 2\%.
From these estimates a total systematic uncertainty of 6--10\% is obtained. 

\begin{figure*}[bt]
\begin{center}
\includegraphics[width=0.5\textwidth]{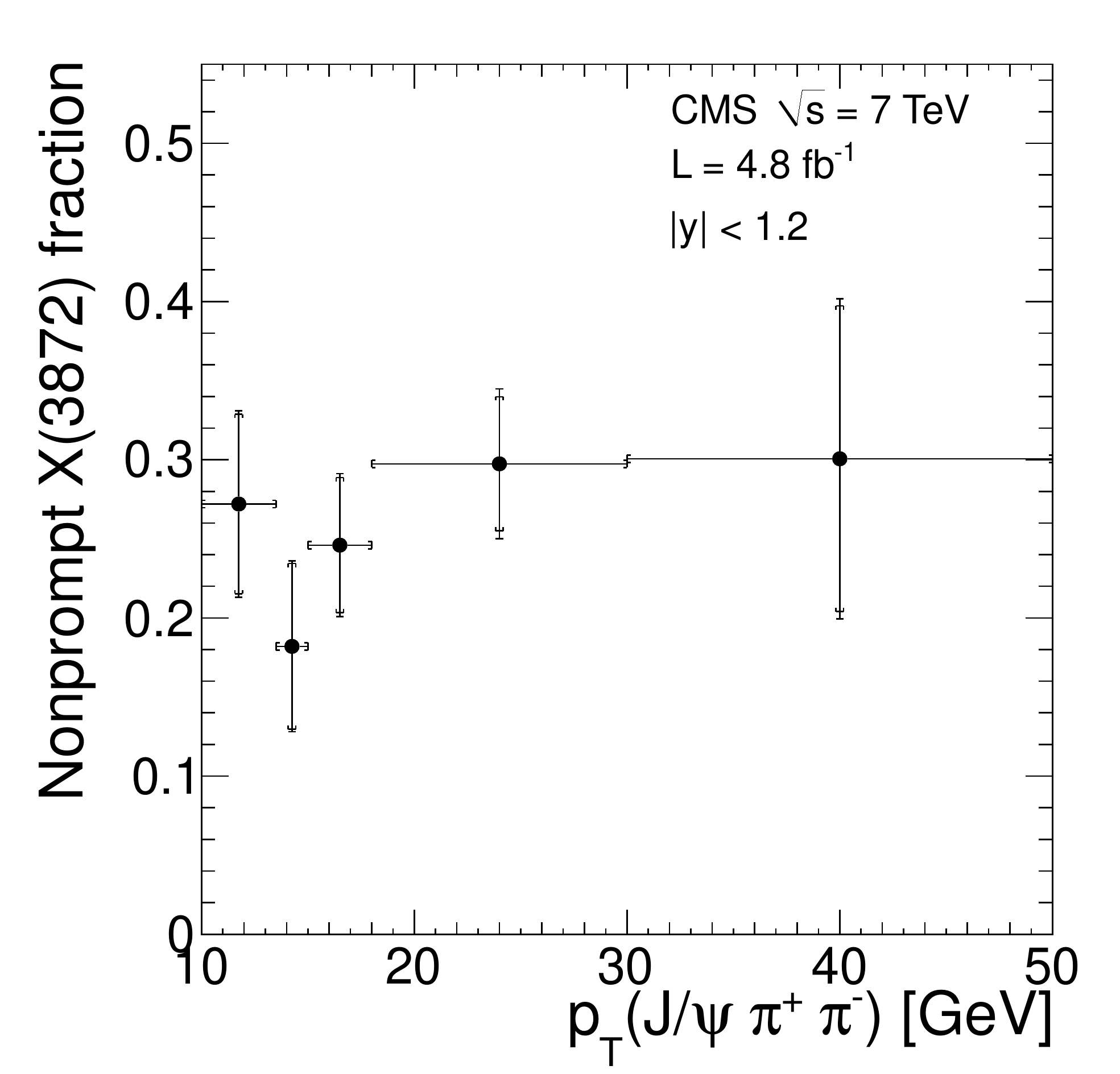}
\caption{Measured $\Xnew$ nonprompt fraction, uncorrected for acceptance, as a function of \pt.
The inner error bars indicate the statistical uncertainty and the outer error bars represent the total uncertainty. The data points are placed at the centre of each \pt bin.
}
\label{fig:NPXvspt}
\end{center}
\end{figure*}

The final results are listed in Table~\ref{tab:BFracResults} and shown in Fig.~\ref{fig:NPXvspt} as a function of \pt.
 The $\Xnew$ nonprompt fraction reveals no significant dependence on transverse momentum and the integrated value
 is significantly smaller than that for the $\psitwo$~\cite{bib-jpsi}.
The results are obtained under the assumption that effects related to the $\Xnew$ polarization cancel in the nonprompt fraction measurement, and therefore no systematic uncertainty is assigned for polarization effects.
\begin{table}[tb]
\topcaption{The $\Xnew$ nonprompt fractions, not corrected for acceptance, as a function of the transverse momentum, together with their statistical and systematic uncertainties, respectively. 
}
\begin{center}
\small
\begin{tabular}{|l|c|c|}
\hline
 Dataset & \pt (\GeVns{}) & $\Xnew$ nonprompt fraction \\
\hline
2011a   & 10--13.5   & 0.272 $\pm$      0.057 $\pm$      0.016   \\
2011a+b & 13.5--15   & 0.182 $\pm$      0.052 $\pm$      0.013   \\
2011a+b & 15--18   & 0.246 $\pm$      0.043 $\pm$      0.015   \\
2011a+b & 18--30   & 0.297 $\pm$      0.042 $\pm$      0.021   \\
2011a+b & 30--50   & 0.301 $\pm$      0.097 $\pm$      0.030   \\  \hline
2011a+b & 10--50   & 0.263 $\pm$ 0.023 $\pm$ 0.016  \\
\hline
\end{tabular}
\end{center}
\label{tab:BFracResults}
\end{table}

\section{Determination of the prompt X(3872) production cross section}
\label{sec:promptxsec}

The cross section times branching fraction for prompt $\Xnew$ production is determined from the measurement of the cross section ratio and the nonprompt fraction, described above, combined with a previous result of the prompt $\psitwo$ cross section~\cite{bib-jpsi}. The latter measurement was performed using the $\psitwo\rightarrow\mu^+ \mu^- $ decay mode and provides results as a function of transverse momentum up to 30\GeV and for the rapidity range $|y|<1.2$.
The prompt $\Xnew$ cross section times branching fraction into $\JPsi \Pgpp \Pgpm$ is given by

\begin{multline*}
\sigma^\text{prompt}_{\Xnew}\cdot\mathcal{B}(\Xnew\rightarrow \JPsi \Pgpp \Pgpm) =\\
 \frac{1 - f^{B}_{\Xnew}}{1 - f^{B}_{\psitwo}} \cdot R \cdot \left( \sigma^\text{prompt}_{\psitwo} \cdot \mathcal{B}(\psitwo\rightarrow\mu^+\mu^-)\right) \cdot \frac{ \mathcal{B}(\psitwo \rightarrow \JPsi \Pgpp \Pgpm)}{\mathcal{B}(\psitwo\rightarrow\mu^+\mu^-)},
\end{multline*}

where $ \sigma^\text{prompt}_{\psitwo} \cdot \mathcal{B}(\psitwo\rightarrow\mu^+\mu^-)$ is the measured prompt $\psitwo$ cross section times $\psitwo\rightarrow\mu^+ \mu^- $ branching fraction~\cite{bib-jpsi}, $R$ is the cross section ratio reported in Section~\ref{sec:inclusive} , and $f^{B}_{\Xnew}$ and $f^{B}_{\psitwo}$ are the nonprompt fractions for $\Xnew$ and $\psitwo$, respectively.
In the calculation, the branching fraction $\mathcal{B}(\psitwo \rightarrow \JPsi \Pgpp \Pgpm)$ is taken from Ref.~\cite{bib-pdg}, and $\mathcal{B}(\psitwo \rightarrow \mu^+ \mu^-)$ is taken to be equal to the more precisely known $\mathcal{B}(\psitwo \rightarrow e^+ e^-)$~\cite{bib-pdg}.

\begin{figure*}[tp]
\begin{center}
\includegraphics[width=0.48\textwidth]{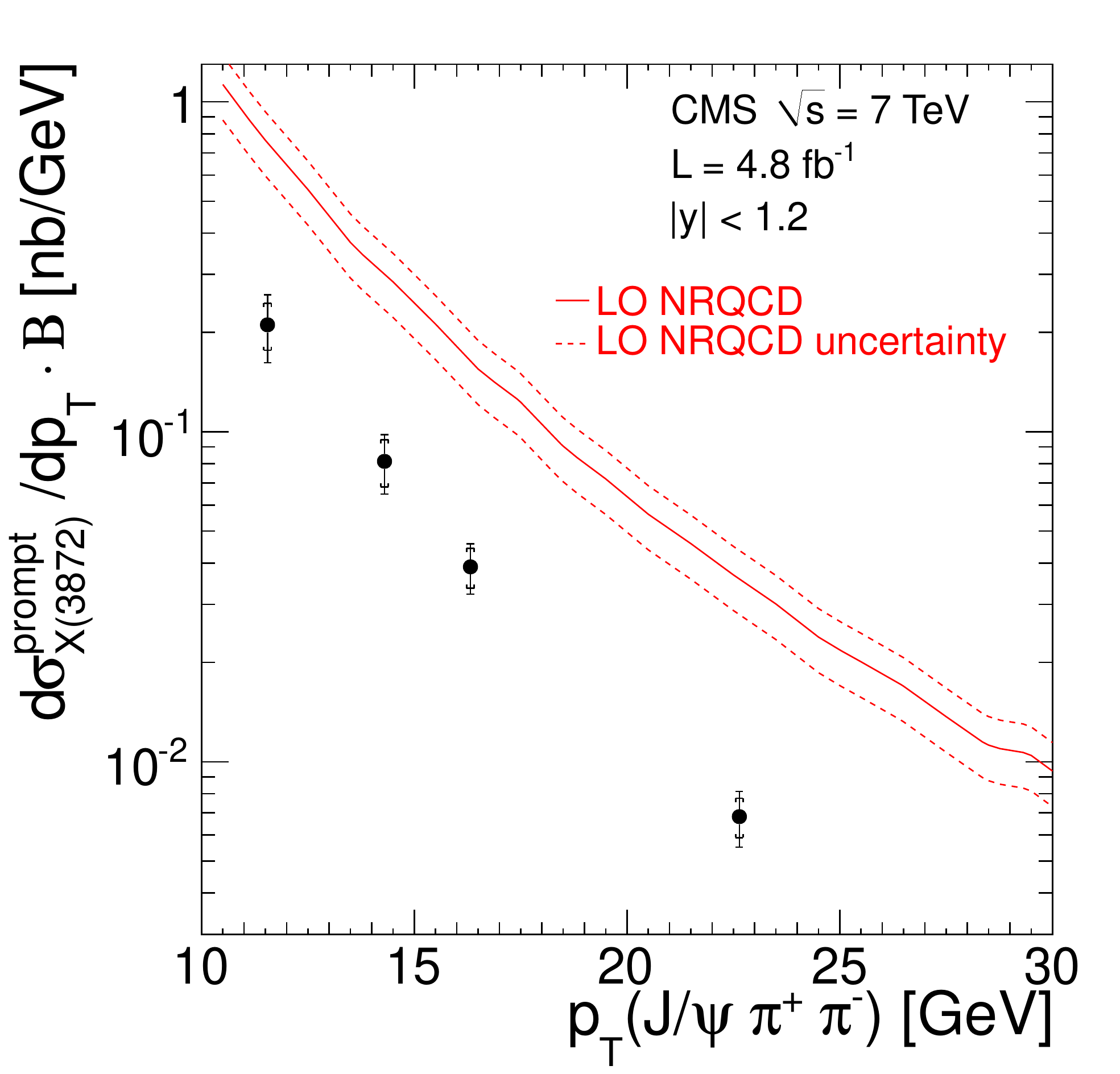}
\caption{Measured differential cross section for prompt $\Xnew$ production times branching fraction $\mathcal{B}(\Xnew \rightarrow \JPsi \Pgpp \Pgpm)$ as a function of \pt. The inner error bars indicate the statistical uncertainty and the outer error bars represent the total uncertainty. Predictions from a NRQCD model~\cite{bib-Artoisenet:2009wk} are shown by the solid line, with the dotted lines representing the uncertainty. The data points are placed where the value of the theoretical prediction is equal to its mean value over each bin, according to the prescription in~\cite{bib-LW}.
}
\label{fig:Promptxsec}
\end{center}
\end{figure*}

The corresponding differential cross section for prompt $\Xnew$ production times the branching fraction to $\JPsi \Pgpp \Pgpm$ as a function of transverse momentum, in the rapidity region $|y| < 1.2$, is listed in Table~\ref{tab:Promptxsec} and shown in Fig.~\ref{fig:Promptxsec}. No cancellation of systematic uncertainties is assumed in the combination. The main sources of systematic uncertainty are related to the measurement of the ratio $R$ and the background lifetime fit in the measurement of the prompt $\psitwo$ cross section~\cite{bib-jpsi}.
\begin{table}[b]
\centering
\topcaption[]{\label{tab:Promptxsec} Prompt $\Xnew$ differential cross section times branching fraction  $\mathcal{B}(\Xnew \rightarrow \JPsi \Pgpp \Pgpm)$ as a function of transverse momentum of the $\JPsi \Pgpp\Pgpm$ system. The uncertainties shown are statistical and systematic, respectively.}
{\small
\begin{tabular}{|c|c|}
\hline
 \pt (GeV) & $\rd\sigma^\text{prompt}_{\Xnew}/\rd\pt \cdot \mathcal{B}(\Xnew \rightarrow \JPsi \Pgpp \Pgpm)$  (nb/\GeVns{})\\
\hline
 10--13.5  & $0.211\phantom{0} \pm 0.034\phantom{0} \pm 0.035\phantom{0}$ \\
 13.5--15  & $0.081\phantom{0} \pm 0.013\phantom{0} \pm 0.010\phantom{0}$ \\
 15--18  & $0.0390 \pm 0.0054 \pm 0.0042$ \\
 15--18  & $0.0390 \pm 0.0054 \pm 0.0042$ \\
 18--30  & $0.0068 \pm 0.0009 \pm 0.0009$ \\
\hline
\end{tabular}
}
\end{table}
A calculation of the predicted differential cross section for prompt $\Xnew$ production in pp collisions at $\sqrt{s} = 7\TeV$ has been made using the NRQCD factorization formalism, assuming the $\Xnew$ is formed from a $\mathrm{c}\overline{\mathrm{c}}$ pair with negligible relative momentum~\cite{bib-Artoisenet:2009wk}.
 This calculation is normalized using Tevatron measurements~\cite{bib-cdf-longlived,bib-cdf-psi2S} with the statistical uncertainty obtained from the experimental input data. The predictions from Ref.~\cite{bib-Artoisenet:2009wk} were modified by the authors to match the phase-space of the measurement presented in this paper.
Comparisons of this prediction with the data, in Fig.~\ref{fig:Promptxsec}, demonstrates that, while the shape is reasonably well described, the predicted cross section is much larger than observed in data.

The integrated prompt $\Xnew$ cross section times branching fraction for the kinematic region $10 < \pt < 30$\GeV and $|y|<1.2$ is also determined. In this kinematic region, the ratio of cross section times branching fraction for $\Xnew$ and $\psitwo$ is $R= 0.0682 \pm 0.0032\stat \pm 0.0065\syst$, and the nonprompt $\Xnew$ fraction is $0.260 \pm 0.024\stat \pm 0.016\syst$. From these results, the measured integrated cross section for prompt $\Xnew$ production times branching fraction 
is:

\begin{equation*}
\label{eq:intxsec}
 \sigma^\text{prompt}(\mathrm{pp} \rightarrow \Xnew + \text{anything}) \cdot \mathcal{B}(\Xnew \rightarrow \JPsi \Pgpp\Pgpm) = 1.06 \pm 0.11\stat \pm 0.15\syst~\text{nb.}
\end{equation*}

This result assumes that the $\Xnew$ and $\psitwo$ states are unpolarized.
The NRQCD prediction for the prompt $\Xnew$ cross section times branching fraction in the kinematic region of this analysis is $4.01 \pm 0.88\unit{nb}$~\cite{bib-Artoisenet:2009wk}, significantly above the measured value.

\section{Measurement of the \texorpdfstring{$\Pgpp\Pgpm$}{pi+ pi-} invariant-mass distribution\label{sec:dipionmass}}

\begin{figure*}[t]
\begin{center}
\includegraphics[width=0.46\textwidth]{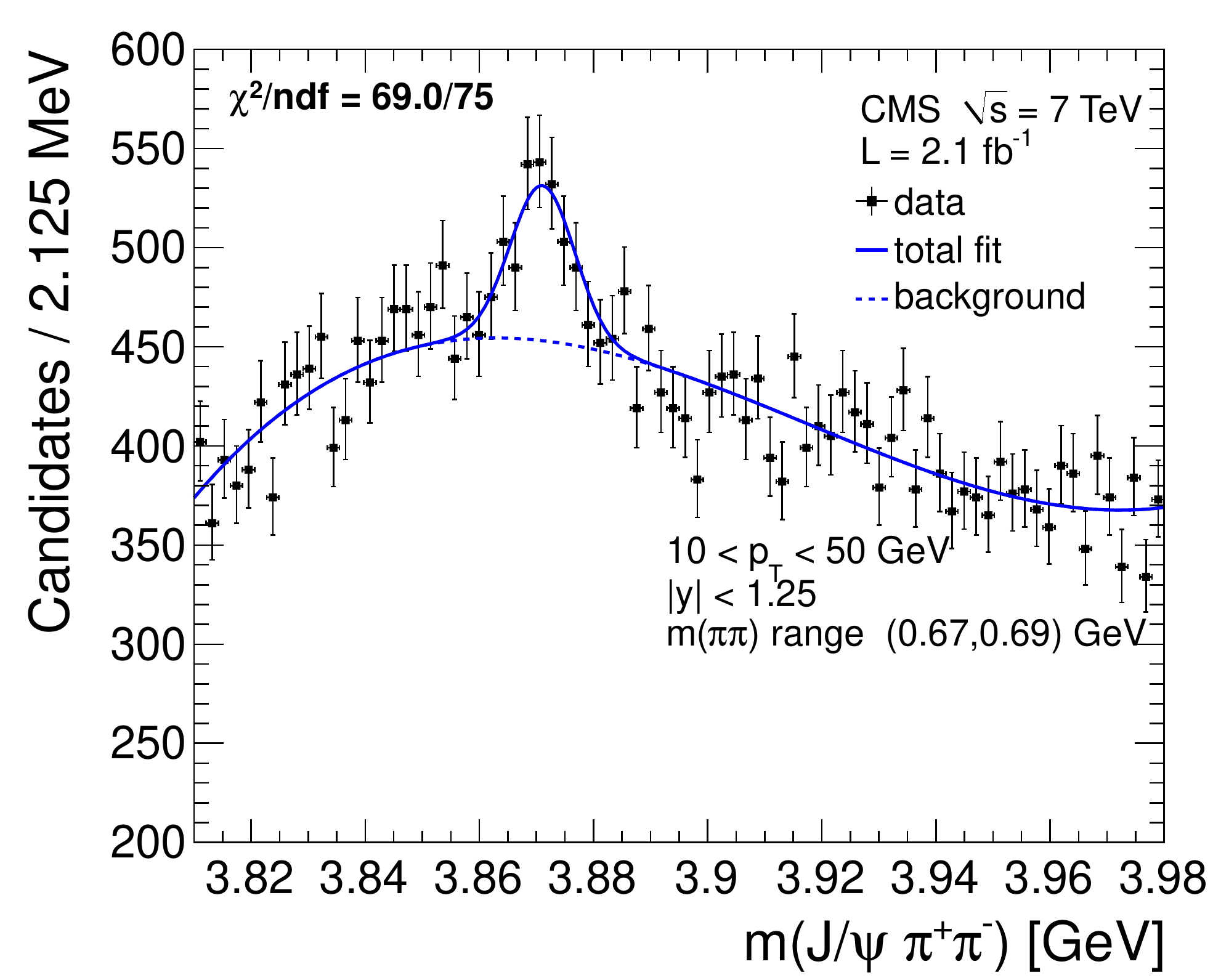}
\includegraphics[width=0.48\textwidth]{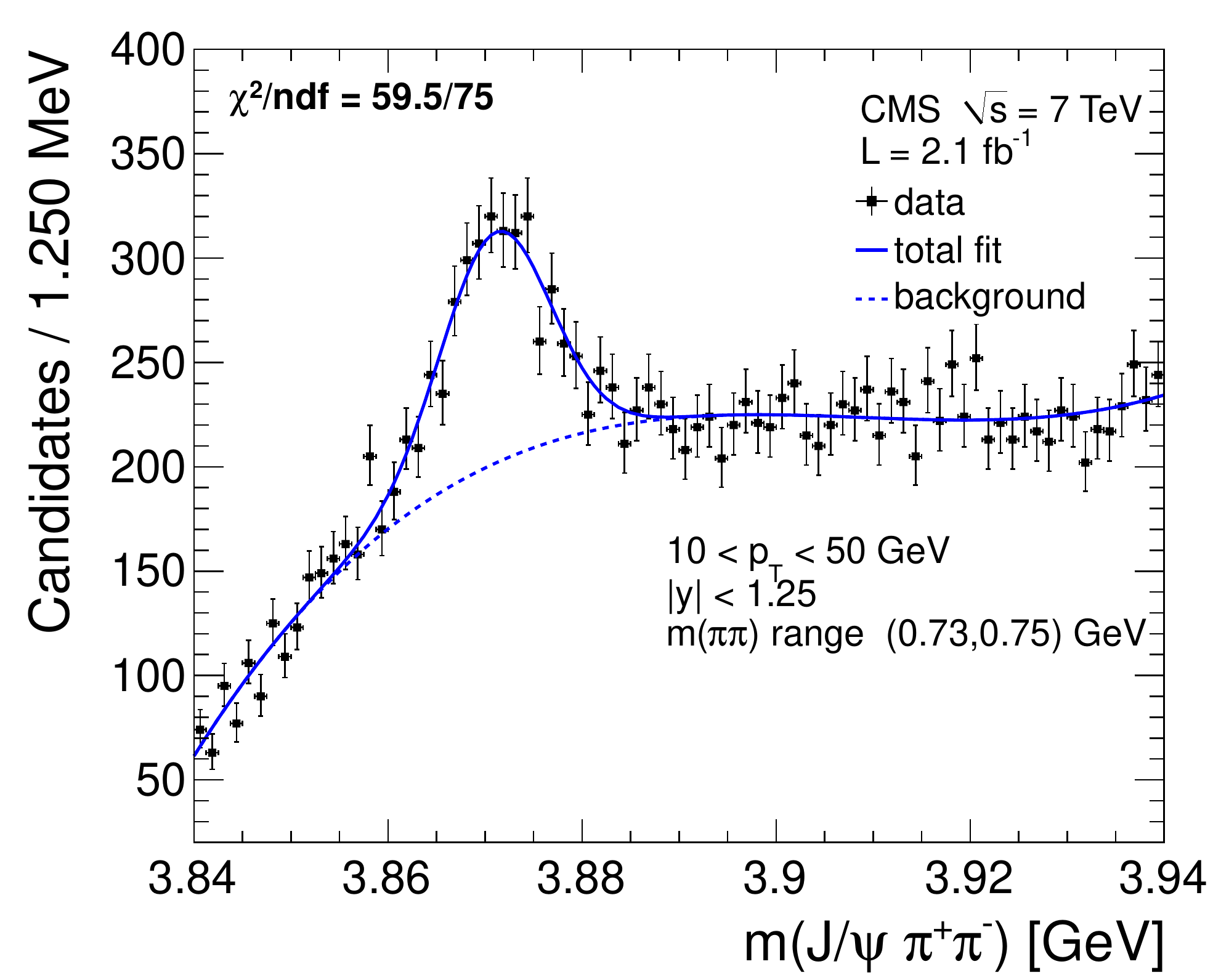}
\caption{Examples of the $\JPsi \Pgpp\Pgpm$ invariant-mass spectrum for the dipion invariant-mass intervals 0.67--0.69\GeV (left) and 0.73--0.75\GeV (right). The lines represent the signal-plus-background fit (solid) and the background-only component (dashed).}
\label{fig:slices}
\end{center}
\end{figure*}
The decay properties of the $\Xnew$ are further investigated with a measurement of the $\Pgpp\Pgpm$ invariant-mass distribution from $\Xnew$ decays to $\JPsi\Pgpp\Pgpm$.
Here, the same event selection as described in Section~\ref{sec:sel} is applied. The event sample 2011a is used, with a transverse momentum threshold of~7\GeV for the muon pair, within the kinematic range $10<\pt<50$\GeV and $|y|<1.25$ for the $\JPsi \Pgpp\Pgpm$.
\begin{figure*}[htb]
\begin{center}
\includegraphics[width=0.48\textwidth]{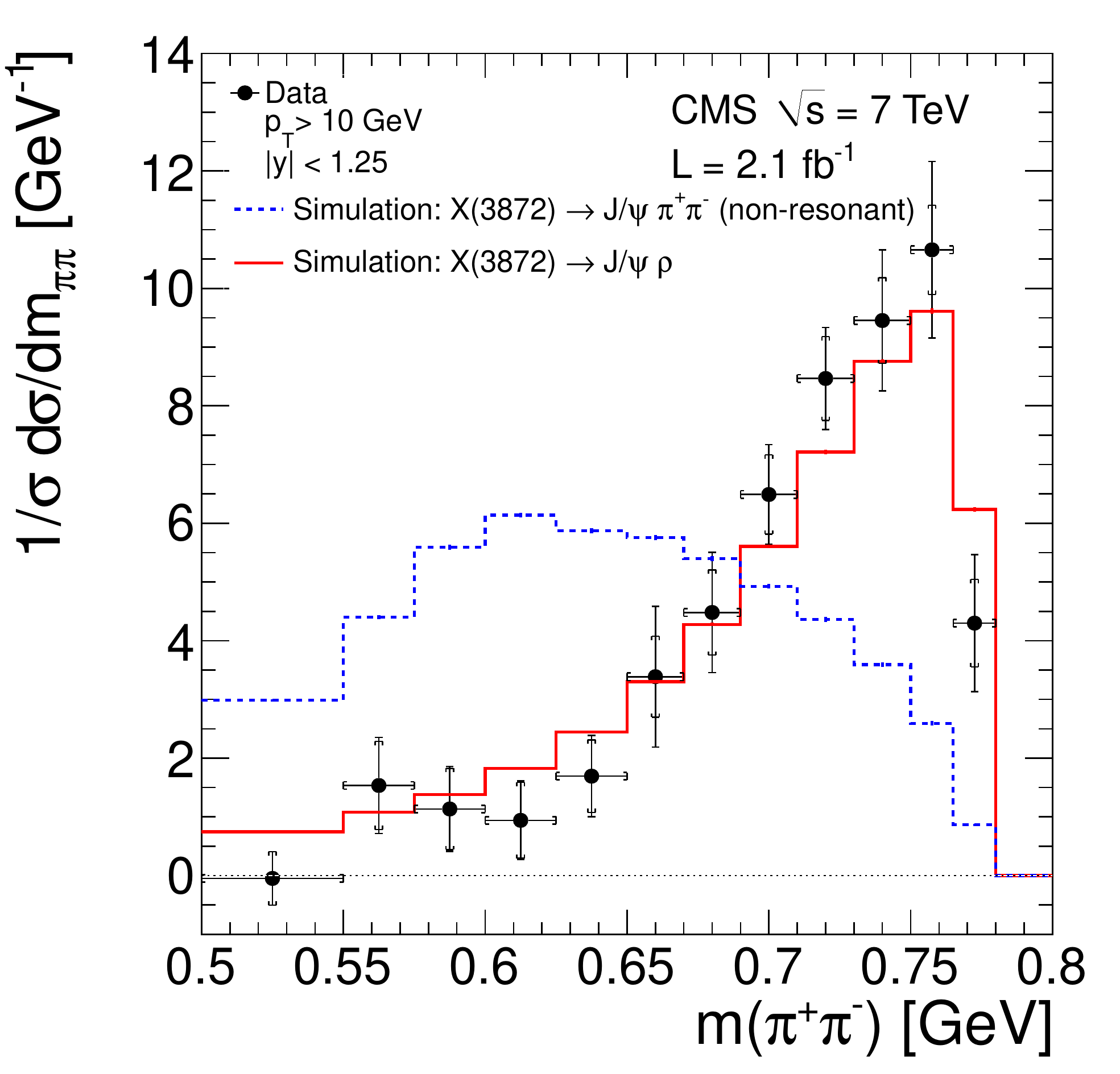}
\caption{
Dipion invariant-mass spectrum for $\Xnew\rightarrow J/\psi \Pgpp\Pgpm$ decays corrected for acceptance and efficiency. The distribution is normalized to unity by dividing by the total cross section for $0.5<m(\Pgpp\Pgpm)<0.78$\GeV. The inner error bars indicate the statistical uncertainty and the outer error bars represent the total uncertainty. The results are compared to results from \EVTGEN with (solid line) and without (dotted line) an intermediate $\rho^0$ decay.
}
\label{fig:masspipi}
\end{center}
\end{figure*}
In this sample, the $\Xnew$ yield with the $\Pgpp\Pgpm$ invariant mass larger than 0.5\GeV is determined from a fit to the $\JPsi \Pgpp\Pgpm$ invariant-mass spectrum to be $6302 \pm 346$, where the uncertainty is statistical only.
The $m(\Pgpp\Pgpm)>0.5$\GeV criterion is imposed to remove events with low efficiency owing to the requirement on the Q value of the decay.

To extract the dipion invariant-mass spectrum from $\Xnew$ decays, the event sample is divided into twelve intervals of dipion invariant mass in the range $0.5 < m(\Pgpp\Pgpm)< 0.78\GeV$. In each interval, a maximum-likelihood fit to the $\JPsi \Pgpp\Pgpm$ invariant-mass distribution is performed, where the signal is modelled with a single Gaussian. The position and width of the $\Xnew$ signal are fixed to the values obtained in the fit to the full sample, except for the last interval, 0.765--0.78\GeV, where the mean and width of the Gaussian are left free to accommodate possible distortions of the signal shape near the upper kinematic limit.
The background shape in $m(\Pgpp\Pgpm)$ intervals is different from the one for the entire $m(\Pgpp\Pgpm)$ spectrum, and a third-order Chebyshev polynomial is used to model it, with the parameters left free in the fit.
The $\JPsi \Pgpp\Pgpm$ invariant-mass spectra for two of the  $\Pgpp\Pgpm$ invariant-mass intervals are shown in Fig.~\ref{fig:slices}.

The $\Xnew$ dipion invariant-mass distribution is extracted from the signal yields obtained from the fits to the data in each interval, after correction for detector acceptance and efficiencies, as estimated from the simulation.
The resulting dipion invariant-mass spectrum, normalized to the total cross section in the interval $0.5<m(\Pgpp\Pgpm)<0.78$\GeV, is presented in Fig.~\ref{fig:masspipi}. 
The data are compared to $\Xnew$ signal simulations with and without an intermediate $\rho^0$ in the $\JPsi\Pgpp\Pgpm$ decay (generation details are described in Section~\ref{sec:sel}). The assumption of an intermediate $\rho^0$ decay gives better agreement with the data, confirming previous measurements~\cite{bib-CDFangluar:2005,bib-Belle:2011}.

Detailed studies are performed to determine the systematic uncertainties.
Scenarios with and without an intermediate $\rho^0$ provide acceptance and efficiency corrections that are very similar.
The impact on the acceptance correction from uncertainties in the $\Xnew$ transverse-momentum spectrum
is found by varying the simulated \pt spectra and generated \pt distribution to match the data. Variations of the corrected yields by 4--6\% are observed and considered as a systematic uncertainty.

The fits to the invariant-mass distributions are done with both free and fixed $\Xnew$ mass and width. In addition, for modelling of the background in the higher dipion invariant-mass bins, a convolution of an exponential and an error functions is used, with a turn-on value constrained to be close to the kinematic limit for each $m(\Pgpp\Pgpm)$ bin.
These variations yield maximal variations of the yields by 10--20\%, and constitute the dominant systematic uncertainty in the measurement of the dipion invariant-mass distribution.

\section{Summary\label{sec:conclusion}}
The $\Xnew$ production cross section has been measured in pp collision at $\sqrt{s}=7$\TeV, with data collected by the CMS experiment, corresponding to an integrated luminosity of 4.8\fbinv. The measurement makes use of the decays of the $\Xnew$ and $\psitwo$ states into $\JPsi \Pgpp \Pgpm$, with subsequent decay of the $\JPsi$ into two muons.
The ratio of the inclusive cross section times branching fraction of the $\Xnew$ and $\psitwo$ in the kinematic region $10<\pt<50$\GeV and $|y|<1.2$ is $R = 0.0656 \pm 0.0029\stat \pm 0.0065\syst$. When restricted to the measured phase-space of the muon and pion pairs, the ratio is $R_\text{fiducial} = 0.0694 \pm 0.0029\stat \pm 0.0036 \syst$. These ratios show no significant dependence on the transverse momentum of the $\JPsi \Pgpp \Pgpm$ system. The results have been obtained with the assumption that the $\Xnew$ has quantum numbers $J^{PC}=1^{++}$ and that both the $\Xnew$ and the $\psitwo$ are unpolarized. Variations of the results for different polarization assumptions have also been reported.
The fraction of $\Xnew$ originating from B-hadron decays is $0.263 \pm 0.023\stat \pm 0.016\syst$,  again assuming the  $\Xnew$ is unpolarized.
No significant dependence on transverse momentum is found. 
From these measurements, the cross section for prompt $\Xnew$ production times branching fraction into $\JPsi \Pgpp \Pgpm$ has been extracted, using a previous CMS measurement of the cross section for prompt $\psitwo$ production. A value of $ \sigma^\text{prompt}(\Pp\Pp \rightarrow \Xnew + \text{anything}) \cdot \mathcal{B}(\Xnew \rightarrow \JPsi \Pgpp\Pgpm) = 1.06 \pm 0.11\stat \pm 0.15\syst\unit{nb}$ is found for the kinematic range $10 < \pt <30$\GeV and $|y|<1.2$.
This result is also made under the assumption that the $\Xnew$ and $\psitwo$ states are unpolarized.
The NRQCD predictions for prompt $\Xnew$ production at the LHC significantly exceed the measured value, while the \pt dependence is reasonably well described.
The measured dipion mass spectrum for $\Xnew \rightarrow\JPsi \Pgpp\Pgpm$ clearly favours the presence of an intermediate $\rho^{0}$ state.

\section*{Acknowledgements}
We would like to thank Pierre Artoisenet and Eric Braaten for modifying their theoretical predictions~\cite{bib-Artoisenet:2009wk} to match the phase-space of our measurement.

\hyphenation{Bundes-ministerium Forschungs-gemeinschaft Forschungs-zentren} We congratulate our colleagues in the CERN accelerator departments for the excellent performance of the LHC and thank the technical and administrative staffs at CERN and at other CMS institutes for their contributions to the success of the CMS effort. In addition, we gratefully acknowledge the computing centres and personnel of the Worldwide LHC Computing Grid for delivering so effectively the computing infrastructure essential to our analyses. Finally, we acknowledge the enduring support for the construction and operation of the LHC and the CMS detector provided by the following funding agencies: the Austrian Federal Ministry of Science and Research and the Austrian Science Fund; the Belgian Fonds de la Recherche Scientifique, and Fonds voor Wetenschappelijk Onderzoek; the Brazilian Funding Agencies (CNPq, CAPES, FAPERJ, and FAPESP); the Bulgarian Ministry of Education, Youth and Science; CERN; the Chinese Academy of Sciences, Ministry of Science and Technology, and National Natural Science Foundation of China; the Colombian Funding Agency (COLCIENCIAS); the Croatian Ministry of Science, Education and Sport; the Research Promotion Foundation, Cyprus; the Ministry of Education and Research, Recurrent financing contract SF0690030s09 and European Regional Development Fund, Estonia; the Academy of Finland, Finnish Ministry of Education and Culture, and Helsinki Institute of Physics; the Institut National de Physique Nucl\'eaire et de Physique des Particules~/~CNRS, and Commissariat \`a l'\'Energie Atomique et aux \'Energies Alternatives~/~CEA, France; the Bundesministerium f\"ur Bildung und Forschung, Deutsche Forschungsgemeinschaft, and Helmholtz-Gemeinschaft Deutscher Forschungszentren, Germany; the General Secretariat for Research and Technology, Greece; the National Scientific Research Foundation, and National Office for Research and Technology, Hungary; the Department of Atomic Energy and the Department of Science and Technology, India; the Institute for Studies in Theoretical Physics and Mathematics, Iran; the Science Foundation, Ireland; the Istituto Nazionale di Fisica Nucleare, Italy; the Korean Ministry of Education, Science and Technology and the World Class University program of NRF, Republic of Korea; the Lithuanian Academy of Sciences; the Mexican Funding Agencies (CINVESTAV, CONACYT, SEP, and UASLP-FAI); the Ministry of Science and Innovation, New Zealand; the Pakistan Atomic Energy Commission; the Ministry of Science and Higher Education and the National Science Centre, Poland; the Funda\c{c}\~ao para a Ci\^encia e a Tecnologia, Portugal; JINR (Armenia, Belarus, Georgia, Ukraine, Uzbekistan); the Ministry of Education and Science of the Russian Federation, the Federal Agency of Atomic Energy of the Russian Federation, Russian Academy of Sciences, and the Russian Foundation for Basic Research; the Ministry of Science and Technological Development of Serbia; the Secretar\'{\i}a de Estado de Investigaci\'on, Desarrollo e Innovaci\'on and Programa Consolider-Ingenio 2010, Spain; the Swiss Funding Agencies (ETH Board, ETH Zurich, PSI, SNF, UniZH, Canton Zurich, and SER); the National Science Council, Taipei; the Thailand Center of Excellence in Physics, the Institute for the Promotion of Teaching Science and Technology of Thailand and the National Science and Technology Development Agency of Thailand; the Scientific and Technical Research Council of Turkey, and Turkish Atomic Energy Authority; the Science and Technology Facilities Council, UK; the US Department of Energy, and the US National Science Foundation.

Individuals have received support from the Marie-Curie programme and the European Research Council (European Union); the Leventis Foundation; the A. P. Sloan Foundation; the Alexander von Humboldt Foundation; the Belgian Federal Science Policy Office; the Fonds pour la Formation \`a la Recherche dans l'Industrie et dans l'Agriculture (FRIA-Belgium); the Agentschap voor Innovatie door Wetenschap en Technologie (IWT-Belgium); the Ministry of Education, Youth and Sports (MEYS) of Czech Republic; the Council of Science and Industrial Research, India; the Compagnia di San Paolo (Torino); and the HOMING PLUS programme of Foundation for Polish Science, cofinanced from European Union, Regional Development Fund.

\bibliography{auto_generated}   

\cleardoublepage \appendix\section{The CMS Collaboration \label{app:collab}}\begin{sloppypar}\hyphenpenalty=5000\widowpenalty=500\clubpenalty=5000\input{BPH-11-011-authorlist.tex}\end{sloppypar}
\end{document}

%% file: BPH-11-011-authorlist.tex
\textbf{Yerevan Physics Institute,  Yerevan,  Armenia}\\*[0pt]
S.~Chatrchyan, V.~Khachatryan, A.M.~Sirunyan, A.~Tumasyan
\vskip\cmsinstskip
\textbf{Institut f\"{u}r Hochenergiephysik der OeAW,  Wien,  Austria}\\*[0pt]
W.~Adam, E.~Aguilo, T.~Bergauer, M.~Dragicevic, J.~Er\"{o}, C.~Fabjan\cmsAuthorMark{1}, M.~Friedl, R.~Fr\"{u}hwirth\cmsAuthorMark{1}, V.M.~Ghete, N.~H\"{o}rmann, J.~Hrubec, M.~Jeitler\cmsAuthorMark{1}, W.~Kiesenhofer, V.~Kn\"{u}nz, M.~Krammer\cmsAuthorMark{1}, I.~Kr\"{a}tschmer, D.~Liko, I.~Mikulec, M.~Pernicka$^{\textrm{\dag}}$, D.~Rabady\cmsAuthorMark{2}, B.~Rahbaran, C.~Rohringer, H.~Rohringer, R.~Sch\"{o}fbeck, J.~Strauss, A.~Taurok, W.~Waltenberger, C.-E.~Wulz\cmsAuthorMark{1}
% ...
\vskip\cmsinstskip
\dag:~Deceased\\
1:~~Also at Vienna University of Technology, Vienna, Austria\\
2:~~Also at CERN, European Organization for Nuclear Research, Geneva, Switzerland\\
% ...
